\documentstyle[aps,eqsecnum,epsfig]{revtex}

\begin{document}

\draft
\preprint{IFUP-TH 2/97}
\draft
\title{Perturbation theory predictions and Monte Carlo simulations for the
2-$d$ $O(n)$ non-linear $\sigma$-models}
\author{B. All\'es, A. Buonanno and G. Cella}
\address{Dipartimento di Fisica dell'Universit\`a and INFN, 
Piazza Torricelli 2, 56126-Pisa, Italy}
\maketitle
\begin{abstract}
By using the results of a high-statistics 
(${\cal O}(10^7)$ measurements) Monte Carlo simulation we
test several predictions of perturbation theory on the $O(n)$
non-linear $\sigma$-model in 2 dimensions. We study the $O(3)$ and
$O(8)$ models on large enough lattices to have a good control 
on finite-size effects.
The magnetic susceptibility and 
three different definitions of the correlation length are
measured. We check our
results with large-$n$ expansions as well as with standard formulae
for asymptotic freedom up to 4 loops in the standard and effective schemes.
 For this purpose the weak coupling expansions of the energy up to 
4 loops for the standard action and up to 3 loops
for the Symanzik action are calculated. For the $O(3)$ model we have used two
different effective schemes and checked that they lead to
compatible results. A great improvement in the results is obtained 
by using the effective scheme based on the energy at 3 and 4 loops.
We find that the $O(8)$ model follows very 
nicely (within few per mille) the perturbative predictions. For
the $O(3)$ model an acceptable agreement (within few per cent) is
found.
\end{abstract}

\vskip 5mm

\section{Introduction}

According to perturbation theory $(PT)$, 
the $O(n)$ non-linear $\sigma$-model
in 2 dimensions for $n \geq 3$ resembles Yang-Mills theories in 4
dimensions. Both are asymptotically free \cite{poly,brez} 
and present a spontaneous
generation of mass. Moreover for $n$=3 the model has a non-trivial
topological content \cite{bela}. Consequently these models are
considered as good toy models for testing methods and solutions in
4-dimensional Yang-Mills theories.
In condensed matter physics these
models have applications in the study of ferromagnetic systems.

There is an extensive literature devoted to investigate the
validity of $PT$ in these models on the lattice 
and in particular the onset of
scaling (see for instance \cite{berg,hase,apos,wolf3,alle}).
In \cite{berg,hase} the $O(3)$ model was analyzed by using
improved actions. 
The obtained results still differ from the exact calculated 
mass-gap \cite{hase2,hase2bis} by 
$\sim 15\%$. In \cite{hase} the authors made use of the 
perturbative $\beta$ function up to 3 loops \cite{falc}.
In \cite{apos,wolf3} faster updating algorithms were used.
The mass-gap for the $O(3)$ model was calculated in \cite{apos}
by using the standard action and an overrelaxed algorithm.
Up to a correlation length $\sim 300$ (in units of lattice spacings) 
it showed a
deviation from the exact result \cite{hase2} of about 20\%.
The $O(4)$ and $O(8)$ models with standard action were studied in
\cite{wolf3} by using the cluster algorithm \cite{wolf}.
The deviation from the exact result for the $O(8)$ model 
at correlation lengths $\sim
30$ was a few per cent. In \cite{alle} an analysis of the performance 
of different lattice geometries for the standard action of the
$O(3)$ model was presented. There was no clear signal
of an earlier onset of asymptotic scaling.

The use of $PT$ for such models is not
guaranteed. The Mermin-Wagner theorem \cite{merm} states 
that continuous 
symmetries in 2-dimensional theories cannot be spontaneously broken. 
Therefore $PT$, which is an expansion around a trivial vacuum, is not 
a priori well-founded.
Motivated by this observation and by the lack of clear asymptotic
scaling in the previous literature, it has been argued \cite{patr} 
that all $O(n \geq 2)$ models undergo a Kosterlitz-Thouless
$(KT)$ \cite{kost,kost2} phase transition at some finite beta $\beta_{KT}$.

In the present work we have performed a high-statistics simulation 
(${\cal O}(10^7)$ measurements) for
the $O(3)$ and $O(8)$ models on the lattice up to correlations
$\sim 130$ for the $O(3)$ model and $\sim 70$ for the $O(8)$ model.
 For the $O(3)$ model we have used the tree-level improved
Symanzik action \cite{syma} and for the $O(8)$ model the standard action.
We have measured 
the magnetic susceptibility and three
different definitions of the correlation length and compared the
results with both the $PT$ and $KT$ set of predictions. 
We have computed also some scaling ratios
which are particularly sensitive to the $PT$ versus $KT$ scenarios.
We have made use of the corrections to asymptotic scaling in $PT$ up to
4 loops in both the standard and effective schemes \cite{mart}
for the $O(8)$ model and up to 3 loops for the $O(3)$ model. 
An effective scheme can be defined by using any short distance dominated
operator; we have used the density of energy operator \cite{mart}.
Hereafter we will call it indistinctly effective or energy scheme.
To include the analysis in this energy scheme, new analytic results are
reported in this paper: the 4-loop coefficient in the weak coupling
expansion of the energy for the standard action and the complete
calculation of all coefficients up to 3 loops for the Symanzik action.

We have used two different definitions of energy operators for the
Symanzik action and checked that the corresponding effective schemes agree. 
Lacking a rigorous treatment for these schemes, this check becomes
an important test.

We have avoided strong coupling effects 
by starting the simulations at large 
enough correlation lengths. The minimal correlation was $\sim 10$ for 
the standard action and $\sim 16$ for the tree-level Symanzik action.

We have not made use of finite-size scaling (questioned
due to the validity of $PT$ whenever the limit 
$\rho \equiv L/\xi \rightarrow 0$
holds, where $L$ is the lattice size and $\xi$ any 
characteristic correlation length)
and we have used rather large lattices ($\rho \approx 7-10$) in order
to control the finite-size effects. We have checked that the 
finite-size effects at these $\rho$ values are negligible.

We are able also to compare the large-$n$ predictions with our
data. In particular, we have checked the relationship between the two
correlation lengths $\xi^{\rm exp}$ and $\xi^{(2)}$ (see
eq. (\ref{eq:defxi}) below) known up to ${\cal O}(1/n)$ and the
prediction for the magnetic susceptibility, known up to ${\cal
O}(1/n^2)$.

In section 2 we will show the predictions of both $PT$ and $KT$ for the
model as well as some necessary $1/n$ expansions. 
In section 3 we will describe our simulations and give the
results while in section 4 we will compare them with the two different
scenarios described in section 2.
In this section we will also use the Monte Carlo data of
ref. \cite{apos} for the $O(3)$ model with standard action to check
the presently known 4-loop perturbative computations.
Our conclusions are given in section 5. 
In the appendix we will 
show some technical details concerning the perturbative
computation of the energy up to 4 loops for the standard action and up
to 3 loops for the Symanzik action.

\section{Predicted scenarios for the $\sigma$-models}

The $O(n)$ non-linear $\sigma$-model in 2 dimensions is defined
formally in the continuum by the action
\begin{equation}
S = {\beta \over 2} \int {\rm d}^2x (\partial_{\mu} {\vec \phi})^2 ,
\end{equation}
where ${\vec \phi}(x)$ is an $n$-component real scalar field,
together with the constraint ${\vec \phi}(x)^2 =1$ for all $x$.
$\beta$ is the inverse of the bare coupling constant. On the lattice
one can regularize this theory by making use of different
actions. For our simulation we chose the standard action 
\begin{equation}
S^{\rm standard} = - \beta \sum_{x,\mu} {\vec \phi}(x) \cdot
{\vec \phi}(x+{\hat \mu})
\label{eq:standard}
\end{equation}
and the tree-level improved Symanzik action \cite{syma}
\begin{equation}
S^{\rm Symanzik} = - \beta \sum_{x,\mu} 
\left( {4 \over 3} {\vec \phi}(x)
\cdot {\vec \phi}(x+{\hat \mu})  - {1 \over {12}} 
{\vec \phi}(x) \cdot {\vec \phi}(x+2{\hat \mu}) \right).
\label{eq:symanzik}
\end{equation}

We have measured the magnetic susceptibility $\chi$ defined as
the zero momentum correlation function,
\begin{equation}
\chi \equiv \sum_{x_1,x_2} G(x_1, x_2), \qquad \qquad
\qquad 
G(x_1,x_2) \equiv \langle {\vec \phi}(0,0) \cdot {\vec \phi}(x_1,x_2)
\rangle 
\label{eq:defchi}
\end{equation}
where we have assumed a symmetric lattice of size $L$ with periodic
boundary conditions in both directions and called $x_1$
and $x_2$ the two coordinates of the point $x$. 
We will need also ${\cal F}$ defined
as the correlation function at the smallest lattice non-zero 
momentum $2 \pi/L$,
\begin{equation}
{\cal F} \equiv \left( {1 \over 2} \sum_{x_1,x_2}
{\rm e}^{2 \pi i x_1/L} G(x_1,x_2) + {1 \over 2} \sum_{x_1,x_2}
{\rm e}^{2 \pi i x_2/L} G(x_1,x_2) \right).
\label{eq:deff}
\end{equation}
We have made use also of the wall-wall correlation function defined as
\begin{equation}
{\bar G}(x_1) \equiv {1 \over L} \sum_{x_2} G(x_1, x_2).
\label{eq:wall}
\end{equation}

We have considered three definitions of correlation lengths,
the exponential one $\xi^{\rm exp}$ and the second momenta of the
correlation function $\xi^{(2)}$ and $\xi'^{(2)}$. They are defined as
$(|x| \equiv \sqrt{x_1^2 + x_2^2})$
\begin{eqnarray}
\xi^{\rm exp} &\equiv&  \displaystyle\lim_{|x| \to \infty}{{-|x|} \over
{\ln G(x_1,x_2)}}, \nonumber \\
\xi^{(2)} &\equiv& { \sqrt{\chi/{\cal F} - 1} \over {2 \sin \pi/L}},
\nonumber \\
\xi'^{(2)} &\equiv& \sqrt{{1\over 4} {{\sum'
 |x|^2 G(x_1,x_2)} \over 
{\sum G(x_1,x_2)}}},
\label{eq:defxi}
\end{eqnarray}
where $\sum'$ indicates that the sum runs over $-L/2+1 \leq x_1,x_2 \leq
L/2$. 
The operative definition of $\xi^{\rm exp}$ on a finite lattice 
was the solution of the equation 
\begin{equation}
{\bar G}(t_1) \cosh \Big( \left( t_2-L/2\right)/
\xi^{\rm exp}\Big) = {\bar G}(t_2) \cosh \Big( \left( t_1-L/2\right)/
\xi^{\rm exp}\Big),
\label{eq:xiexpeq}
\end{equation}
for big enough $t_1$ and $t_2$
where ${\bar G}(t)$ is the wall-wall correlation 
and $t_2 - t_1 = \Delta_t$ with $\Delta_t=1$, 2.
As a function of $t_1$, the solution of the previous
equation displays a long stable plateau for 
$\xi^{\rm exp} \lesssim t_1 \lesssim 3 \xi^{\rm exp}$. 
Anyhow, we chose the value and error
for $\xi^{\rm exp}$ self-consistently at $t_1 \approx 2 \xi^{\rm exp}$.
The result is independent of $\Delta_t$ (both for the Symanzik action
and the standard one) and we selected the value
$\Delta_t=1$. In Figure 1 we show an example of solution of
eq. (\ref{eq:xiexpeq}) as a function of $t_1$; the plateau is apparent.

On the other hand, the value for the definition $\xi'^{(2)}$ 
was extracted from the wall-wall correlation function
\begin{equation}
\xi'^{(2)} = \sqrt{ {1 \over 2} 
{{\sum'
 t^2 {\bar G}(t)} \over 
{\sum {\bar G}(t)}}}.
\end{equation}
In the large-$L$ limit $\xi^{(2)}$ and $\xi'^{(2)}$ coincide. For
finite $L$ the three definitions show rather different finite-size
behaviours \cite{lusccom,cara1}.

\begin{figure}[htbp]
\epsfig{file=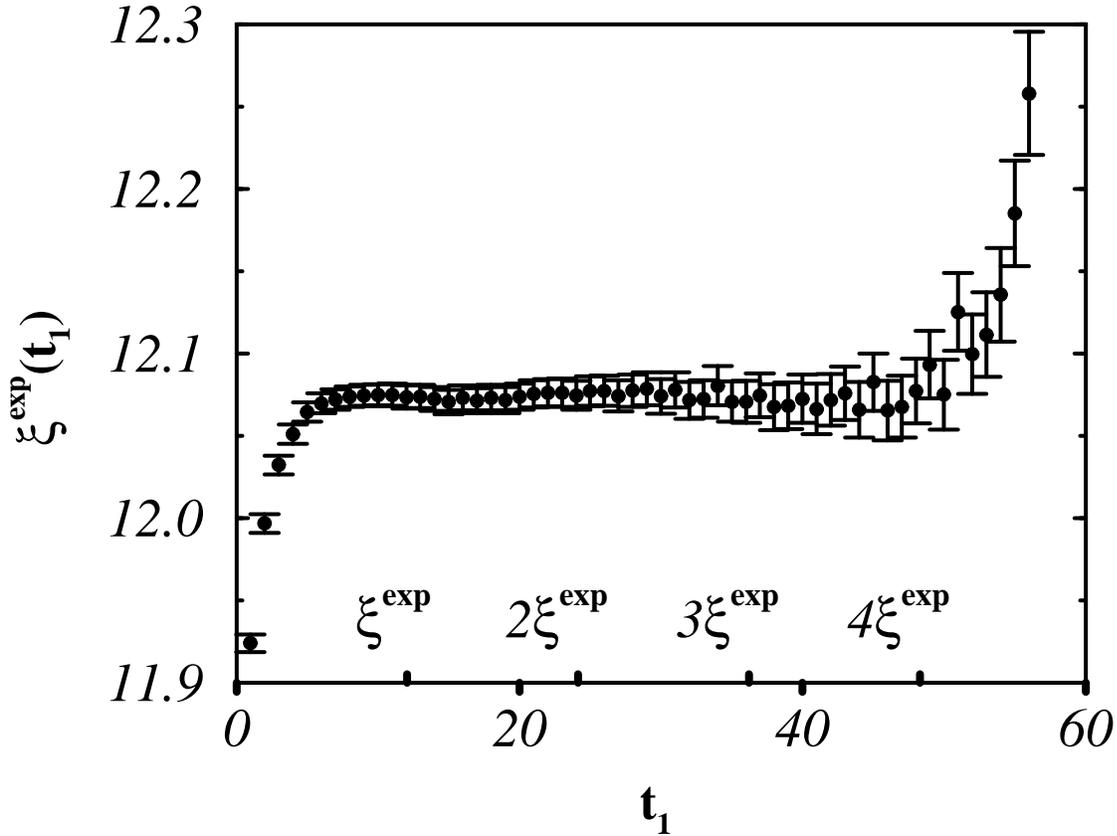,width=0.9\textwidth}
\caption{
Solution of eq. (\ref{eq:xiexpeq}) as a function of
$t_1$ for $\Delta_t=1$ for the $O(8)$ model at 
$\beta=4.8$ and $L=120$. 
The parameter $t_1$ is given in units of lattice spacings (external
labels of the horizontal axis) and units of correlation 
length (internal labels).}
\label{fig:1}
\end{figure}

The scaling of these quantities as predicted in perturbation theory 
in the large-$L$ limit is
\begin{eqnarray}
\xi &=& C_{\xi} \left( {{n - 2} \over {2 \pi \beta}}
\right)^{\displaystyle 1 \over \displaystyle{n-2}} 
\exp \left( {\displaystyle 
{2 \pi \beta} \over \displaystyle {n-2}} \right)
\left( 1 + \sum_{k=1}^{\infty} {a_k \over \beta^k} \right), \nonumber \\
\chi &=&  C_{\chi} \left( {{n - 2} \over {2 \pi \beta}}
\right)^{\displaystyle {n+1} \over \displaystyle{n-2}} 
\exp \left({\displaystyle 
{4 \pi \beta} \over \displaystyle{n-2}} \right)
\left( 1 + \sum_{k=1}^{\infty} {b_k \over \beta^k} \right).
\label{eq:scaling}
\end{eqnarray}
The form of the expression for $\xi$ is valid for all three 
definitions. The coefficients
of non-universal scaling $a_k$ and $b_k$ are action-dependent. They
are known up to 3 loops for the Symanzik action \cite{falc} 
and up to 4 loops for the standard action \cite{cara2}. The
constant $C_{\xi}$ is definition- and action-dependent (its dependence
on the action is exactly calculable in perturbation theory up to an
universal constant). $C_{\xi}$ is known exactly for the
exponential definition.
With the standard action it is \cite{hase2,hase2bis}
\begin{equation}
C_{\xi^{\rm exp}} = \left( {{\rm e} \over 8} 
\right)^{\displaystyle 1 \over \displaystyle {n-2}}
\Gamma\left(1 + {1 \over {n-2}}\right) \; 
2^{-5/2} \; \exp \left( - {{\pi} \over 
{2 (n-2)}} \right).
\label{eq:hasenfratz}
\end{equation}
The corresponding constant in the tree-level 
Symanzik action is easily obtained from (\ref{eq:hasenfratz}) by using 
the exact perturbative result \cite{syma,berg2}
\begin{equation}
{\Lambda_{\rm Symanzik} \over \Lambda_{\rm standard}} =\exp 
\left( {{0.2964 \; n - 0.0920} \over {n-2}} \right).
\label{eq:berg}
\end{equation}
The other constants are not exactly known. For the correlation lengths
in the large-$n$ limit we have~\cite{cara2}
\begin{equation}
C_{\xi'^{(2)}} =
C_{\xi^{(2)}} = C_{\xi^{\rm exp}} \left( 1 - {{0.0032} \over n}
 + {\cal O}\left({1 \over n^2}\right) \right).
\label{eq:constantscorr}
\end{equation}
In the same limit the value of $C_\chi$ with the standard action 
is \cite{bisc}
\begin{equation}
C_{\chi} = 0.196 \left( 1 - {{4.267} \over n} 
+ {\cal O}\left({1 \over n^2}\right) \right).
\label{eq:constantschi}
\end{equation}
In ref. \cite{flyv} there are numerical 
results for $C_{\chi}$ up to ${\cal
O}(1/n^2)$,
\begin{eqnarray}
&O(3)& \; \; \; {\rm (standard \; action)} 
\;\;\longrightarrow \;\; C_\chi = 0.0127 \nonumber \\
&O(8)& \; \; \;{\rm (standard \; action)} \;\;\longrightarrow \;\; C_\chi = 0.103 
\label{eq:flyvresults}
\end{eqnarray}
By using eq. (\ref{eq:berg}) the value of $C_{\chi}$ for the Symanzik
action can be obtained; for $n=3$ it is $C_{\chi}=0.0625$.
 From eq. (\ref{eq:scaling}) we conclude that the ratio 
\begin{equation}
R_{PT} \equiv {\chi \over \xi^2} \left( {{2 \pi \beta} \over {n-2}}
\right)^{\displaystyle {n-1} \over \displaystyle{n-2}} 
\left( 1 + {\cal O}\left({1 \over{\beta}} \right) \right)
\label{eq:ptratio}
\end{equation}
tends to the constant $C_{\chi}/C_{\xi}^2$ as the continuum 
limit $\beta \rightarrow \infty$ is approached. 
The parentheses contain the corrections which are known up to 4
loops for the standard action and 3 loops for the Symanzik action.
Hereafter we will call this ratio the $PT$ ratio.

The perturbative expansions of the energy for both the standard action
and the Symanzik action are calculated in the appendix.

The correlation length for the $O(2)$ model, when $\tau \equiv
\beta_{KT}-\beta$ is positive and small, scales as \cite{kost,kost2}
\begin{equation}
\xi = A \exp \left( { B \over \tau^{1/2}} \right),
\label{eq:ktxi}
\end{equation}
with $A$ and $B$ positive constants. On the other hand the ratio
\begin{equation}
R_{KT} \equiv {{\chi \; \tau^{r}} \over \xi^{2 -\eta}}
\label{eq:ktratio}
\end{equation}
should be constant as we approach $\beta_{KT}$ from below. Here 
$\eta=1/4$ is the critical exponent. Following the renormalization group
considerations of ref. \cite{kost,kost2} 
one can show that $r=1/16$~\cite{grins}.
Recent numerical analyses for the $O(2)$ model 
\cite{bute,kenn,patr2,jank,campo} have yielded several values 
for $r$ which are all consistent with the bound $|r| \lesssim 0.1$.
Eqs. (\ref{eq:ktxi}-\ref{eq:ktratio}) are the expected behaviour
(and consistent with 
Monte Carlo simulations) for the $O(2)$ model. 
 From now on we will call the ratio in eq. (\ref{eq:ktratio}) 
the $KT$ ratio.
The $KT$ scenario for the $O(n)$ model is the extension of
this behaviour for $n \geq 3$.

In ref. \cite{patr2} a fit of Monte Carlo 
data for $\chi$ and $\xi$ for the 
$O(3)$ model with standard action was performed. Within errors 
it gave a constant for the $KT$ ratio and a strong decrease 
far from constant for the $PT$ ratio.
We have simulated the $O(3)$ model with the tree-level Symanzik action
\cite{syma} in order to check the results of ref. \cite{patr2} with an
action classically closer to the continuum limit.

\section{The Monte Carlo simulation}

We have performed an extensive Monte Carlo simulation of the $O(3)$
model with Symanzik action and the $O(8)$ model with standard action.
In tables I and II we show the corresponding sets of raw data.
The statistical error of the three entries for 
$\xi'^{(2)}$, $\xi^{(2)}$ in table II
for $L=50,100,200$ and the two entries for $L=150,300$ in table I 
display a strong dependence 
on the lattice size. This can be explained by
taking into account that the definition of, for example $\xi'^{(2)}$,
involves the quantity $\sum |x|^2 G(x)$. For a big enough lattice the
``signal'' is independent on size, while the ``noise'' grows as the
volume. A similar argument can be used for $\xi^{(2)}$.

\vskip 1cm

{\centerline {\bf Table I}}
{\it Raw Monte Carlo data for $O(3)$ with Symanzik action.
The second row for $L=300$ was used only for checks of finite-size
dependence.}

\vskip 5mm

\moveright 0.8 in
\vbox{\offinterlineskip
\halign{\strut
\vrule \hfil\quad $#$ \hfil \quad & 
\vrule \hfil\quad $#$ \hfil \quad & 
\vrule \hfil\quad $#$ \hfil \quad & 
\vrule \hfil\quad $#$ \hfil \quad & 
\vrule \hfil\quad $#$ \hfil \quad & 
\vrule \hfil\quad $#$ \hfil \quad & 
\vrule \hfil\quad $#$ \hfil \quad \vrule \cr
\noalign{\hrule}
\beta & L & 10^{-6} \cdot {\rm stat} & \chi & \xi^{\rm exp} & 
\xi^{(2)} & \xi'^{(2)} \cr
\noalign{\hrule}
1.40 & 150 & 8.1 & 361.41(26) & 16.216(23) & 16.195(17) & 15.441(17) \cr
\noalign{\hrule}
1.40 & 300 & 8.08 & 360.99(27) & 16.168(27) & 16.161(71) & 16.050(52) \cr
\noalign{\hrule}
1.45 & 200 & 8 & 587.18(44) & 21.519(30) & 21.443(23) & 20.462(22) \cr
\noalign{\hrule}
1.50 & 260 & 6.24 & 972.78(83) & 28.793(50) & 28.730(35) & 27.273(34) \cr
\noalign{\hrule}
1.55 & 340 & 18 & 1634.41(83) & 38.668(61) & 38.636(36) & 36.410(41) \cr
\noalign{\hrule}
1.60 & 450 & 2.88 & 2777.3(3.6) & 52.72(38) & 52.42(10) & 49.18(25) \cr
\noalign{\hrule}
1.65 & 600 & 4 & 4743.1(5.2) & 71.08(34) & 70.93(13) & 66.27(20) \cr
\noalign{\hrule}
1.70 & 800 & 7.25 & 8125.7(6.7) & 96.67(37) & 96.28(13) & 89.64(19) \cr
\noalign{\hrule}
1.75 & 1050 & 1.24 &  13852.7(27.7) & - & 130.25(36) & - \cr
\noalign{\hrule}
}}

\vskip 5mm

\vskip 1cm

{\centerline {\bf Table II}}
{\it Raw Monte Carlo data for $O(8)$ with standard action.
The first and third rows ($L=50$ and $L=200$ respectively) were used 
only for checks of finite-size dependence.}

\vskip 5mm

\moveright 0.8 in
\vbox{\offinterlineskip
\halign{\strut
\vrule \hfil\quad $#$ \hfil \quad & 
\vrule \hfil\quad $#$ \hfil \quad & 
\vrule \hfil\quad $#$ \hfil \quad & 
\vrule \hfil\quad $#$ \hfil \quad & 
\vrule \hfil\quad $#$ \hfil \quad & 
\vrule \hfil\quad $#$ \hfil \quad & 
\vrule \hfil\quad $#$ \hfil \quad \vrule \cr
\noalign{\hrule}
\beta & L & 10^{-6} \cdot {\rm stat} & \chi & \xi^{\rm exp} & 
\xi^{(2)} & \xi'^{(2)} \cr
\noalign{\hrule}
4.6 & 50 & 16 & 145.501(77) & 9.7787(75) & 9.7541(40) & 7.5678(69) \cr
\noalign{\hrule}
4.6 & 100 & 3.76 & 149.00(17) & 9.881(14) & 9.864(13) & 9.533(20) \cr
\noalign{\hrule}
4.6 & 200 & 16 & 149.029(83) & 9.8624(74) & 9.860(35) & 9.842(14) \cr
\noalign{\hrule}
4.7 & 110 & 16 & 177.86(10) & 10.9156(80) & 10.913(12) & 10.543(13) \cr
\noalign{\hrule}
4.8 & 120 & 16 & 212.13(12) & 12.0745(88) & 12.071(13) & 11.635(15) \cr
\noalign{\hrule}
4.9 & 140 & 16 & 253.11(14) & 13.370(10) & 13.358(17) & 13.005(17) \cr
\noalign{\hrule}
5.0 & 160 & 40 & 302.600(85) & 14.806(10) & 14.788(11) & 14.431(10) \cr
\noalign{\hrule}
5.4 & 220 & 40 & 620.78(18) & 22.193(28) & 22.199(15) & 21.381(15) \cr
\noalign{\hrule}
5.8 & 340 & 65 & 1289.32(30) & 33.526(47) & 33.411(16) & 32.357(21) \cr
\noalign{\hrule}
6.0 & 290 & 3.2 & 1854.6(2.5) & 40.933(81) & 40.879(78) & 36.530(87) \cr
\noalign{\hrule}
6.1 & 320 & 3.2 & 2236.6(3.0) & 45.52(10) & 45.425(87) & 40.49(11) \cr
\noalign{\hrule}
6.2 & 360 & 3.2 & 2689.7(3.7) & 50.21(11) & 50.26(10) & 44.99(12) \cr
\noalign{\hrule}
6.3 & 390 & 3.2 & 3239.7(4.5) & 55.77(12) & 55.60(11) & 49.35(14) \cr
\noalign{\hrule}
6.4 & 440 & 3.2 & 3909.8(5.4) & 61.94(15) & 61.75(12) & 55.06(15) \cr
\noalign{\hrule}
6.5 & 480 & 3.2 & 4686.9(6.5) & 68.36(17) & 68.16(14) & 60.64(16) \cr
\noalign{\hrule}
}}

\vskip 5mm

We have updated the configurations with the Wolff 
algorithm \cite{wolf}. 
We verified the absence of autocorrelations in the data 
for the standard action. For the Symanzik action we have 
used a generalization of this algorithm \cite{buon} which does
not completely eliminate the critical slowing down.
According to the measured integrated 
autocorrelation time \cite{buon}, we have
performed 4 decorrelating updatings for this action between successive
measurements. Once these decorrelating updatings were done, we
explicitly checked the absence of autocorrelations in the data
for the Symanzik action. We have 
measured $\chi$ and the three definitions of $\xi$ shown in the
previous section. The necessary two-point correlation functions
were evaluated by using an improved estimator \cite{wolf2}.

\vskip 1cm

{\centerline {\bf Table III}}
{\it Integrated autocorrelation times $\tau^{\rm int}_{1,2}$
for the energies $E^S_{1,2}$ and size $\langle C_\# \rangle$ of the
average Fortuin-Kasteleyn cluster 
for a Symanzik-improved action simulated on a lattice size $L=100$.}

\vskip 5mm

\moveright 1 in
\vbox{\offinterlineskip
\halign{\strut
\vrule \hfil\quad $#$ \hfil \quad & 
\vrule \hfil\quad $#$ \hfil \quad & 
\vrule \hfil\quad $#$ \hfil \quad & 
\vrule \hfil\quad $#$ \hfil \quad & 
\vrule \hfil\quad $#$ \hfil \quad & 
\vrule \hfil\quad $#$ \hfil \quad \vrule \cr
\noalign{\hrule}
\beta & E^S_1 & \tau^{\rm int}_1 & E^S_2 & \tau^{\rm int}_2 & 
\langle C_\# \rangle \cr
\noalign{\hrule}
2. & 0.98252(5) & 23.6(1.6) & 0.7788(2) & 26.4(1.8) & 3464.9 \cr
\noalign{\hrule}
5. & 1.14782(2) & 29.4(2.2) & 0.91534(4) & 31.4(2.4) & 6148.7 \cr
\noalign{\hrule}
10. & 1.19949(2) & 40.4(3.5) & 0.95812(4) & 42.7(3.8) & 7321.1 \cr
\noalign{\hrule}
}}

\vskip 5mm

We have also done a few runs at small physical volumes,
$\rho=L/\xi \ll 1$, to calculate the energy for the Symanzik-improved action 
at large $\beta$, (see appendix). We have 
realised that the performance of the extension of the Wolff algorithm 
for Symanzik actions \cite{buon} 
is less effective in this regime. In Table III we give the 
integrated autocorrelation times $\tau^{\rm int}_{1,2}$
for the calculation of the energies $E^S_1$ and $E^S_2$ respectively 
on a lattice of size $L=100$ after
$7\; 10^5$ measurements for several $\beta$. 
The integrated autocorrelation times must be compared 
with the values $\tau^{\rm int} \sim 4$ found 
when $\rho \gg 1$ \cite{buon}. In table III we also give 
the size $\langle C_\# \rangle$
of the average Fortuin-Kasteleyn cluster \cite{fk,sw}.
At very small physical volumes the result of a single Wolff updating 
is an almost global flip of the entire lattice, thus becoming an
approximate reflection symmetry of the
whole system. From Table III we see that the average cluster size 
becomes larger as $\beta$ increases (the total number of sites in our
lattice is 10000). The worsening of the performance of the algorithm in the 
$\rho\ll 1$ regime can be traced back to this fact. 
Such behaviour is also visible if the standard action Wolff
algorithm is used.

We have run our simulations at very high statistics obtaining rather
small statistical errors. Therefore the systematic 
errors can become relevant and
they require a careful analysis.
We consider three sources of such errors: the finite-size
effects, the different constants in front of the scaling for the
correlation length and the non-universal corrections to asymptotic
scaling. 

All observables, (other than the energy at very high $\beta$), 
have been measured at values
of the ratio $\rho \gtrsim 7$. For the $O(8)$ model and 
$\beta < 6.0$ 
this ratio was $\rho \gtrsim 10$. With these $\rho$ values 
the finite-size effects are few parts per mille and we will not
consider them. We have checked this
assertion by performing a few runs at different values of the previous
ratio. For the $O(3)$ model with Symanzik action at $\beta=1.40$ we have 
used the lattice size $L=150$, 300 ($\rho=9,\;18$ respectively) 
as shown in Table I. The values obtained for $\chi$, $\xi^{\rm exp}$
and $\xi^{(2)}$ are compatible for both sizes. Only  
$\xi'^{(2)}$ shows a clear size dependence. We have imposed the 
predicted $L$ dependence \cite{cara1} obtaining 
$\xi^{(2)}(L) = \xi^{(2)}(\infty) + 3.9(10.6)/\rho^2$ and
$\xi'^{(2)}(L) = \xi'^{(2)}(\infty) - 6.2(7)\; \rho\;\exp(-\rho/2)$.
We see that although the size dependence of $\xi'^{(2)}$ has an 
exponential fall-off \cite{cara1}, the presence of the multiplicative 
$\rho$ and the large coefficient in front of the
exponential makes our data for $\xi'^{(2)}$ at 
$\rho \gtrsim 7$ not reliable enough. 
Instead the data for $\xi^{(2)}$ are good in spite of
the presence of the power-law $1/\rho^2$. 
The size dependence of the data for 
$\xi^{\rm exp}$ is as gentle as for the $\xi^{(2)}$ data.

On the other hand for the $O(8)$ model with standard action 
we have simulated the value $\beta=4.6$ at three lattice sizes:
50, 100 and 200 ($\rho=5$, 10 and 20). Again only $\xi'^{(2)}$
displays clearly a size dependence. 
Fitting the data to an exponential for $\xi'^{(2)}$ and a power-law
for $\xi^{(2)}$ \cite{cara1} we obtain
$\xi^{(2)}(L) = \xi^{(2)}(\infty) - 3.0(10.0)/\rho^2$ and
$\xi'^{(2)}(L) = \xi'^{(2)}(\infty) - 5.7(5) \;\rho\;\exp(-\rho/2)$.
As before, the $L$-dependence is sizeable only for
$\xi'^{(2)}$ due to the large coefficient in front of the 
$\rho$-function. The data for
$\xi^{\rm exp}$ display a size dependence as mild as that for $\xi^{(2)}$.

As for the unknown non-perturbative constant $C_{\xi^{(2)}}$, when
$n=3$ eq. (\ref{eq:constantscorr}) gives 
$C_{\xi^{(2)}}/C_{\xi^{\rm exp}} =0.9989$. The value for this
ratio provided by 
the data in Table I is 0.9979(9). 
In ref. \cite{meyer} the values 0.9994(8) and 0.9991(9) are quoted for
$\beta=1.7$ and $\beta=1.8$ respectively.
This ratio for $n=8$ from 
eq. (\ref{eq:constantscorr}) is 0.9996 and from the data of Table II
is 0.9989(4). We conclude that eq. 
(\ref{eq:constantscorr}) is reliable although the ${\cal O}(1/n^2)$
term would be welcome. 

In our subsequent analysis we will make use of the data for 
$\xi^{(2)}$ in both $O(3)$ and $O(8)$ because this
definition for the correlation is less size-dependent than
$\xi'^{(2)}$ and on the other hand allows
a better error determination than for the exponential definition,
(to evaluate the error of $\xi^{(2)}$ we also measured the
cross correlation between $\chi$ and ${\cal F}$).
We will correct the non-perturbative constant $C_{\xi^{\rm exp}}$,
eq. (\ref{eq:hasenfratz}),
by dividing all data by 0.9979(9) and 0.9989(4) for $O(3)$ and
$O(8)$ respectively.

The corrections to universal
scaling are the largest source of errors and will be discussed in 
the next section.

\section{Discussion of results}

In tables I and II we show the raw data for $\chi$ and the three
definitions of $\xi$. In the following analysis we will use the values
for $\xi^{(2)}$ and we will write $C_\xi$ instead of $C_{\xi^{(2)}}$.
As we said in the previous section we shall neglect
the finite-size effects and introduce 
a corrective factor to the prediction 
(\ref{eq:hasenfratz}) for $C_{\xi}$. 

\begin{figure}[htbp]
\epsfig{file=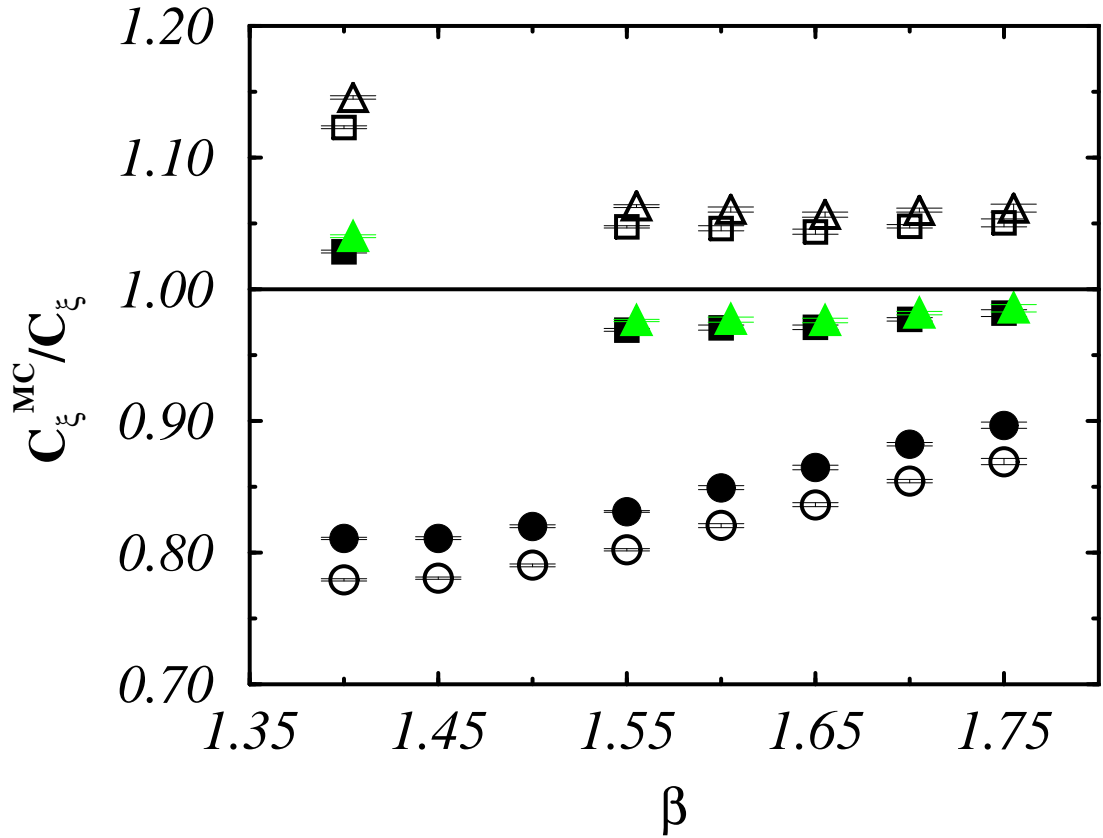,width=0.9\textwidth}
\caption{
The ratio between non-perturbative constants
$C^{\rm MC}_{\xi}/C_{\xi}$ for the $O(3)$ model with Symanzik action.
Circles (squares, triangles) stand for standard scheme ($E_1^S$-scheme,
$E_2^S$-scheme). Open (full) symbols mean 2-loop (3-loop) data.
The data for the $E_2^S$-scheme have been slightly shifted horizontally to
render the figure clearer.}
\label{fig:2}
\end{figure}

\subsection{The $O(3)$ model with Symanzik action}

 From the data for $\xi^{(2)}$ and
eq. (\ref{eq:scaling}) we can compute the constant $C_{\xi}$.
We shall call $C^{\rm MC}_{\xi}$ such constant
obtained from the Monte Carlo data. If $PT$ is correct and asymptotic
scaling holds, this number should be independent of $\beta$ and equal
to the prediction of eq. (\ref{eq:hasenfratz}) for $n$=3. Therefore
the ratio $C^{\rm MC}_{\xi}/C_{\xi}$ should be 1.
In Figure 2 we show
such ratio as a function of $\beta$ by using eq. (\ref{eq:scaling}) at
2-loop and 3-loop \cite{falc} for both the standard and energy schemes. 
We used two different energy schemes defined by the operators
$E_1^S$ and $E_2^S$, (eqs. (\ref{eq:energysy1},\ref{eq:energysy2})
of the appendix). The respective $\beta_E$ are
\begin{equation}
\beta_{E1} \equiv {{w_1^{S1}} \over {15/12 - E_1^S}}, \qquad \qquad
\beta_{E2} \equiv {{w_1^{S2}} \over {1 - E_2^S}}.
\label{eq:betae}
\end{equation}
The perturbative expansions of $E_1^S$ and $E_2^S$ are given in the
appendix and the Monte Carlo values for both operators are listed
in Table IV.

\newpage

{\centerline {\bf Table IV}}
{\it Measured values of the two operators (\ref{eq:energysy1})
and (\ref{eq:energysy2}) for the $O(3)$ model with Symanzik action.}

\vskip 5mm

\moveright 1.2 in
\vbox{\offinterlineskip
\halign{\strut
\vrule \hfil\quad $#$ \hfil \quad & 
\vrule \hfil\quad $#$ \hfil \quad & 
\vrule \hfil\quad $#$ \hfil \quad & 
\vrule \hfil\quad $#$ \hfil \quad & 
\vrule \hfil\quad $#$ \hfil \quad \vrule \cr
\noalign{\hrule}
\beta & L &  10^{-5} \cdot \hbox{stat} &E^S_1 & E^S_2 \cr
\noalign{\hrule}
1.40 & 300 & 4 & 0.840997(3) & 0.661762(2) \cr
\noalign{\hrule}
1.55 & 340 &  4 & 0.890991(2) & 0.703158(1) \cr
\noalign{\hrule}
1.60 & 450 &  4 & 0.904603(1) & 0.714419(1) \cr
\noalign{\hrule}
1.65 & 600 &  4 & 0.917106(1) & 0.724757(1) \cr
\noalign{\hrule}
1.70 & 800 &  4 & 0.928573(1) & 0.734231(1) \cr
\noalign{\hrule}
1.75 & 1050 &  0.36 & 0.93917(2) & 0.74299(1) \cr 
\noalign{\hrule}
5.0 & 100 &  7 & 1.14782(2) & 0.91534(4) \cr
\noalign{\hrule}
10.0 & 100  &  7 & 1.19949(2) & 0.95812(4) \cr
\noalign{\hrule}
15.0 & 100 &  4 & 1.21650(3) & 0.97222(2) \cr
\noalign{\hrule}
}}

\vskip 5mm

 Figure 2 displays an asymptotic
approach to unity for increasing $\beta$. 
The data in the standard scheme (circles) differ from unity by $\sim 15\%$.
This is in accordance with previous
numerical calculations of $C_{\xi}$ with the tree-level Symanzik 
action \cite{hase}. However, the lack of asymptotic
scaling in the energy schemes (squares and triangles) 
amounts only to $2-3\%$ at 3 loops. 
Notice also that the two energy schemes agree fairly well; this
fact supports the reliability of these schemes. This agreement improves
as $\beta$ increases. In the previous section we saw that the 
systematic error in the corrective factor
$C_{\xi^{(2)}}/C_{\xi^{\rm exp}}$ was of the order of 1 per mille which
is negligible in Figure 2.

\begin{figure}[htbp]
\epsfig{file=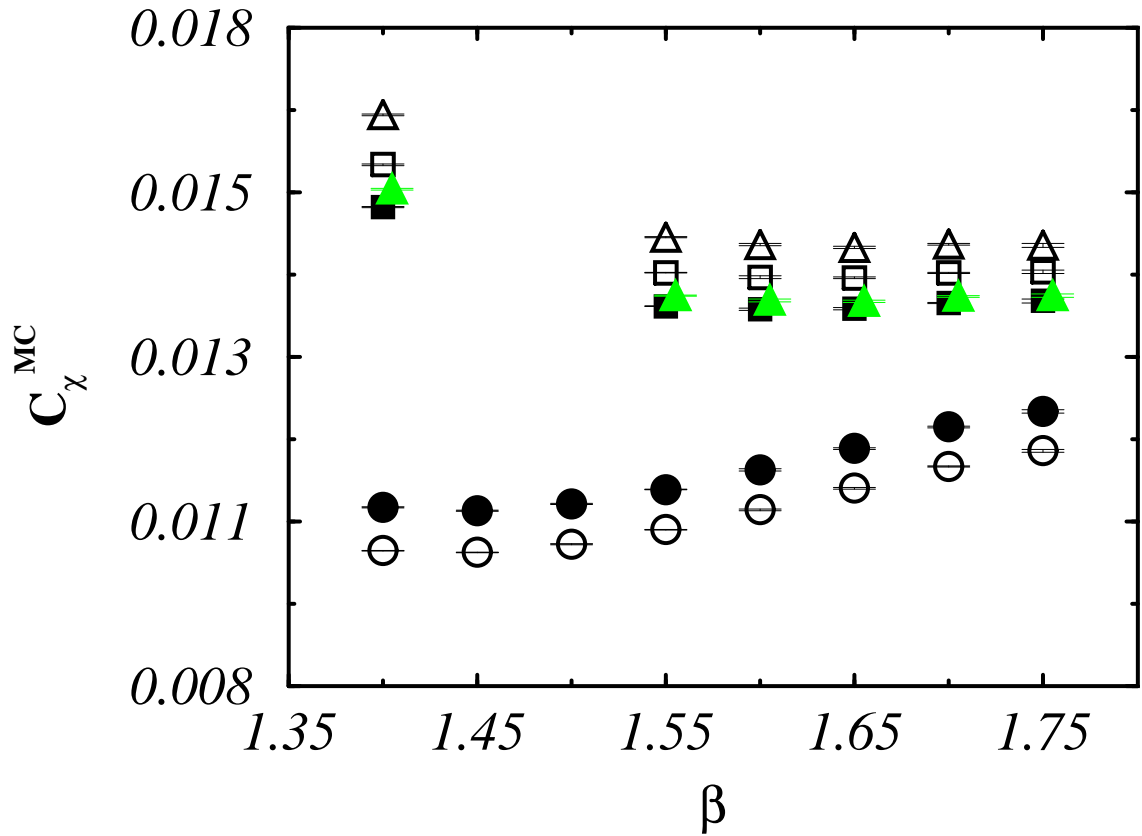,width=0.9\textwidth}
\caption{
The non-perturbative constant $C^{\rm MC}_{\chi}$
for the $O(3)$ model with Symanzik action. The constant is given in
units of $\Lambda_{\rm standard}$.
Circles (squares, triangles) stand for standard scheme ($E_1^S$-scheme,
$E_2^S$-scheme). Open (full) symbols mean 2-loop (3-loop) data.
The data for the $E_2^S$-scheme at 3-loop has been slightly shifted
horizontally to render the figure clearer.}
\label{fig:3}
\end{figure}

\begin{figure}[htbp]
\epsfig{file=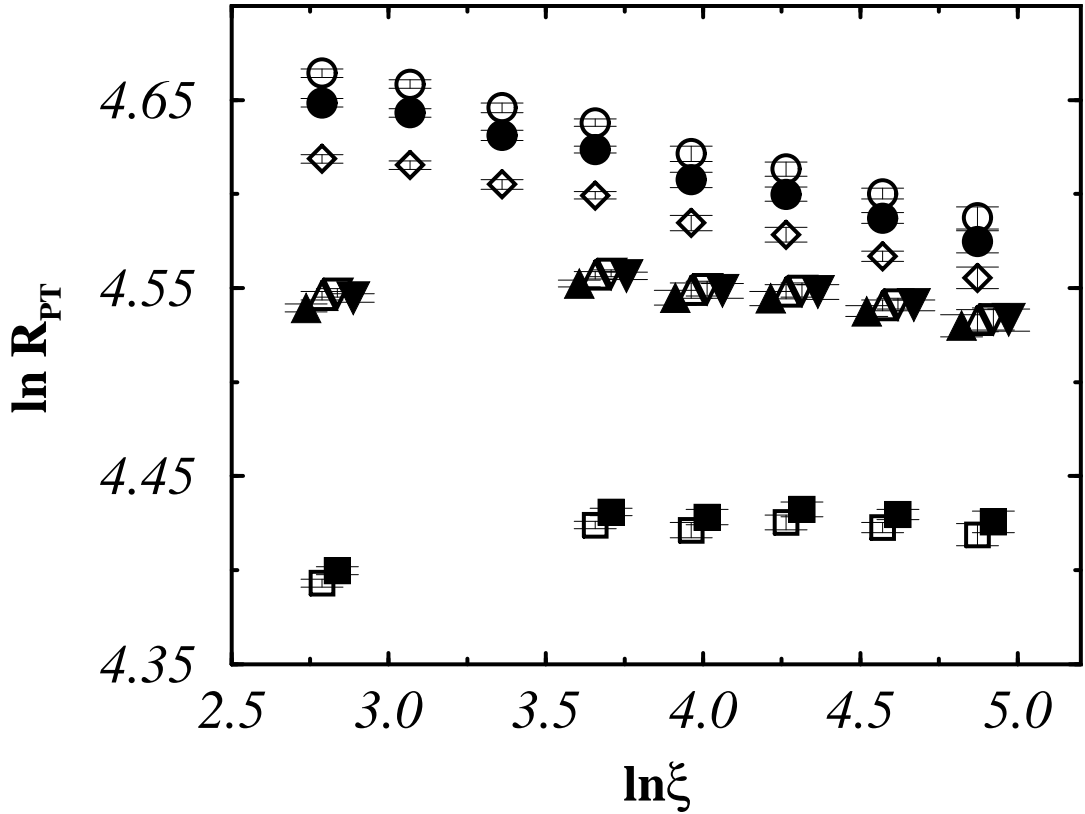,width=0.9\textwidth}
\caption{
The $PT$ ratio (\ref{eq:ptratio}) for the
for the $O(3)$ model with Symanzik action.
Open circles (full circles, open diamonds) stand for 
the 1-loop (2-loop, 3-loop) result in the standard scheme. Open
squares (open up-triangles, open down-triangles) stand for the 
1-loop (2-loop, 3-loop) result in the $E_1^S$-scheme. Full
squares (full up-triangles, full down-triangles) stand for the 
1-loop (2-loop, 3-loop) result in the $E_2^S$-scheme.
Some data have been slightly shifted horizontally to
render the figure clearer.}
\label{fig:4}
\end{figure}

In Figure 3 we show the constant $C_{\chi}$ computed from our Monte
Carlo data at 2 and 3 loops in the standard and energy schemes. 
At present there are no available exact
predictions for this constant. The $1/n^2$ calculation \cite{flyv}
provides~0.0625 for the tree-level Symanzik action. 
After rescaling with (\ref{eq:berg}) this number becomes 0.0127. From 
the 3-loop data in the energy scheme of Figure 3 one
can infer the estimate $C^{\rm MC}_{\chi} =0.0138(2)$ which differs
by $\sim 8\%$ from the large-$n$ calculation. This result can be compared
with the estimate of ref. \cite{cara4} which is 0.0146(11); we see that both
agree within errors (notice that at 2 loops our result would be 0.0145(3);
this error includes the imprecision among the $E^S_1$ and $E^S_2$ data). 
The estimate of ref. \cite{cara4} has been obtained by using finite-size
scaling techniques \cite{fs1,fs2}.

We observe that the $1/n$-expansion up to
order ${\cal O}(1/n^2)$ converges well, hence we expect 
a better agreement if further corrections were added. 
Finally, notice that the two energy schemes agree much better at 3-loop
than at 2-loop.

The results for the $PT$ ratio are reported in Figure 4 up to
3 loops. The data in the standard scheme are far from constant
although, as is known for the
Symanzik-improved actions \cite{berg}, the slope is less steep than for the
standard action case \cite{patr2}. 
The data for the two energy schemes at 2 and 3-loop agree completely.
Moreover these data are flatter indicating that 
scaling has possibly set in. 
Assuming this onset of scaling, we derive from the data at 2 and 3 loops
in these effective schemes $\ln (C_\chi/C_\xi^2) = 4.54(2)$.
Using the prediction (\ref{eq:hasenfratz}) 
$C_\xi = 0.01249$, we
obtain $C_\chi = 0.0146(2)$ in good agreement with
the value inferred from Figure 3.
To show the result at 3-loop, the corresponding coefficient of the gamma
function for the tree-level Symanzik action has been used. This coefficient
can be obtained from \cite{falc} after correcting a misprint in their
eq. (25): the $(2\pi)^4$ dividing the last term in $Z_1$ must be instead
$(2\pi)^2$. We thank M. Falcioni for correspondence about this point,
\cite{private}.
The 3-loop coefficient is thus
\begin{equation}
\gamma_2^{\rm Symanzik} = {1 \over {(2 \pi)^3}} 
\left( -2.01 + 1.65 \; n + 0.362 \; n^2 \right)
\end{equation}

\begin{figure}[htbp]
\epsfig{file=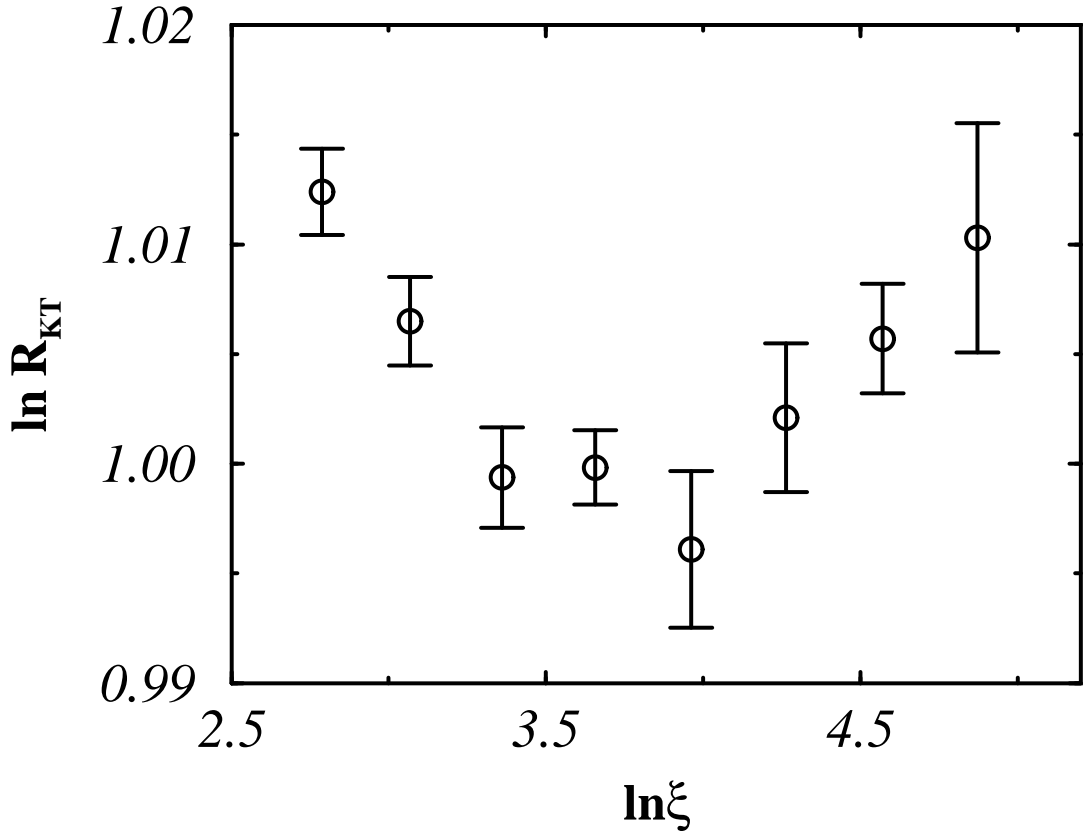,width=0.9\textwidth}
\caption{The $KT$ ratio (\ref{eq:ktratio}) for the 
$O(3)$ model with Symanzik action.}
\label{fig:5}
\end{figure}

Now we want to test the $KT$ formulae
(\ref{eq:ktxi}) and (\ref{eq:ktratio}). A best fit of
the data for $\xi$ to eq. (\ref{eq:ktxi}) is rather unstable.
This can be understood as follows: assuming that the $KT$
transition point $\beta_{KT}$ does exist and it is far away, we
can expand eq. (\ref{eq:ktxi}) in powers of $\beta/\beta_{KT}$
obtaining
\begin{equation}
\xi \approx A \exp\left(B \over \sqrt{\beta_{KT}}\right) 
\exp\left({B\beta} \over {2\beta_{KT}^{3/2}}\right) 
\equiv A' \exp\left(B' \beta\right).
\label{eq:ktxiapprox}
\end{equation}
This equation shows that actually we are fitting the combination
$B'\equiv B/(2\beta_{KT}^{3/2})$, therefore the best fit cannot
yield reliable information about the precise value of $\beta_{KT}$.
However, the fact that the previous analysis within $PT$ gave rather
acceptable results indicates that the linear approximation in 
eq.~(\ref{eq:ktxiapprox}) is good and indeed $\beta_{KT}$ is
much larger than our working $\beta$'s.

In Figure 5 the results for the $KT$ ratio (\ref{eq:ktratio}) 
are shown. By using the previous conclusion about the large value
of $\beta_{KT}$, we have assumed that inside the narrow interval 
$1.4 < \beta < 1.75$ the factor $\tau^{r}$ in 
(\ref{eq:ktratio}) is almost constant. 
As a consequence we did not consider it.
In ref. \cite{patr2} this ratio, calculated for the standard action, 
looked almost constant with the critical
exponent $\eta=1/4$. 
We emphasize that our data 
have smaller error-bars and so the
interval in the vertical axis is almost 7 times finer for our data.
This fact allows us to see that our 
result is clearly not constant. 
We have estimated the probability $Q$ that the data in Figure 5 follow a 
straight line. $Q$ is obtained from the tail of the $\chi^2$ 
probability distribution, (we have assumed a gaussian distribution 
for the point ordinates).
We have obtained less than $Q=0.01$ which means that with probability 
$\sim 99\%$ the data do not follow a straight line.
We have repeated the same analysis after removing the first two points
(one can argue that they are still far from the scaling region of
the $KT$ transition). In this case $Q=0.09$ which still indicates
that the data do not lie on a straight line with probability 91\%.
If the constancy of this ratio was
to be a true physical effect then our data for the Symanzik-improved
action should stay also constant. 

A similar probability calculation shows that also the 2 and 
3-loop data in the 
energy scheme of Figure 4 do not follow a straight line (although
the 1-loop data in this scheme is essentially flat). 
We remark, however, that the effective schemes and the 
loop corrections have flattened out the
data in the $PT$ ratio. In contrast, the increase of the resolution in the 
statistics has revealed that the $KT$ ratio is not as flat as claimed in
ref. \cite{patr2}.

Our results for the $O(3)$ model in the standard scheme do not confirm 
either of the two scenarios. The lack of asymptotic scaling
agrees with previous works using the same Symanzik
improved action, \cite{hase}.
However, in the energy schemes these data
present asymptotic scaling at 3 loops within $2-3\%$ for the
correlation length as well as an estimate for the magnetic susceptibility
that is in reasonable accordance with previous numerical simulations
\cite{cara4} and the $1/n$ expansion. 
The $PT$ ratio in the energy scheme shows a much flatter behaviour
than in the standard scheme.
Moreover the agreement between the two energy schemes is a reassuring result.

On the other hand, in the $KT$ scenario, we have seen that 
the scaling law (\ref{eq:ktratio}) is badly satisfied. This is
in constrast with the data of \cite{patr2} for the standard action. 

\subsection{The $O(3)$ model with standard action}

In Table V we show the Monte Carlo results for the $O(3)$ model
with standard action taken from ref. \cite{apos} and the
Monte Carlo energy, (see eq. (\ref{eq:energyst}) of the appendix), 
from ref. \cite{mendes}.
The correlation length data corresponds to the exponential
definition in eq. (\ref{eq:defxi}), so there is no correction
factor in this case.

\vskip 1cm

{\centerline {\bf Table V}}
{\it Data for the $O(3)$ model with standard action.
The $\chi$ and $\xi^{\rm exp}$ data has been taken
from \cite{apos}; the energy data from \cite{mendes}.}

\vskip 5mm

\moveright 0.9 in
\vbox{\offinterlineskip
\halign{\strut
\vrule \hfil\quad $#$ \hfil \quad & 
\vrule \hfil\quad $#$ \hfil \quad & 
\vrule \hfil\quad $#$ \hfil \quad & 
\vrule \hfil\quad $#$ \hfil \quad & 
\vrule \hfil\quad $#$ \hfil \quad & 
\vrule \hfil\quad $#$ \hfil \quad \vrule \cr
\noalign{\hrule}
\beta & L \;(\hbox{for $\chi$, $\xi^{\rm exp}$}) &  \chi & \xi^{\rm exp} 
& L \;(\hbox{for $E$}) & E\cr
\noalign{\hrule}
1.50 & 256 &  176.4(2) & 11.05(1) & 128 & 0.601597(16) \cr
\noalign{\hrule}
1.60 & 256 &  448.4(7) & 19.00(2) & 128 & 0.635722(10) \cr
\noalign{\hrule}
1.70 & 512 &  1263.7(3.3) & 34.44(6) & 256 & 0.664240(5) \cr
\noalign{\hrule}
1.75 & 768 &  2197.(15.) & 47.2(2) & 256 & 0.676629(4) \cr
\noalign{\hrule}
1.80 & 768 &  3823.(21.) & 64.5(5) & 256 & 0.687953(3) \cr
\noalign{\hrule}
1.85 & 768 & 6732.(25.) & 88.7(5) & 256 & 0.698351(3) \cr
\noalign{\hrule}
1.90 & 1024 & 11867.(62.) & 122.7(1.1) & 256 & 0.707952(3) \cr
\noalign{\hrule}
1.95 & 1024  & 20640.(310.) & 164.8(5.3) & 128 & 0.716928(9) \cr
\noalign{\hrule}
}}

\vskip 5mm

The asymptotic scaling analysis for these data was done up
to 3 loops in 
\cite{apos} while the test for the $KT$ scenario was done in \cite{patr2}.
Here we want to make use of our new perturbative results for the 
energy up to 4 loops, (eq. (\ref{eq:n3}) of the appendix), 
and the results of \cite{cara2} to test asymptotic
scaling in the energy scheme for the magnetic susceptibility, 
the correlation length 
and the $PT$ ratio. The energy scheme is defined as
\begin{equation}
\beta_E \equiv {w_1 \over {1 - E}}.
\end{equation}

In Figure 6 we show the ratio $C^{\rm MC}_{\xi}/C_{\xi}$. The
lack of asymptotic scaling in the standard schemes is apparent
and the energy scheme does not improve it as dramatically as
for the Symanzik action. We see that the 
4-loop correction in the energy scheme
is almost negligible and as a result the departure from asymptotic
scaling at 3-loop observed in \cite{apos} is still present at 4-loop.
The lack of asymptotic scaling in this figure is $\sim 10\%$ for the
energy scheme and $15-20\%$ for the standard one.

 Figure 7 displays the non-perturbative constant $C_{\chi}$ as computed
from the Monte Carlo data. The data in the energy scheme 
converge around $C_\chi =0.0130(5)$ while the $1/n^2$ prediction
\cite{flyv} is 0.0127 and the result of \cite{cara4} was 0.0146(11).
The result with our data for the Symanzik action was 0.0138(2).
The several Monte Carlo results are compatible with each other
suggesting that the truncation error of the series at order $1/n^2$
amounts to $\sim 8\%$ when $n=3$.

\begin{figure}[htbp]
\epsfig{file=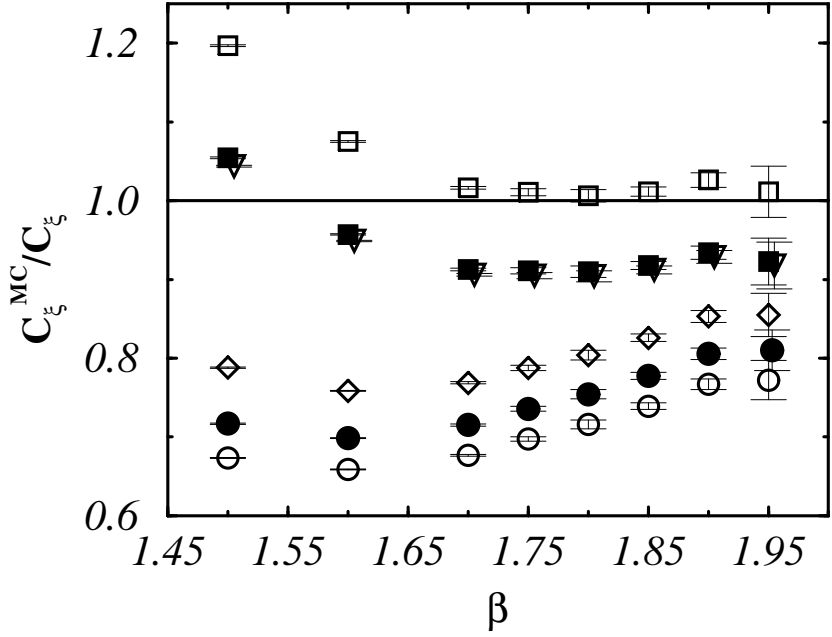,width=0.7\textwidth}
\caption{The ratio between non-perturbative constants
$C^{\rm MC}_{\xi}/C_{\xi}$ for the $O(3)$ model with standard action.
Monte Carlo data from \protect\cite{apos}. Open circles (full circles,
open diamonds) correspond to 2-loop (3-loop, 4-loop) in the
standard scheme; open squares (full squares, open triangles)
correspond to the 2-loop (3-loop, 4-loop) in the energy scheme.
Some data have been slightly shifted horizontally 
to render the figure clearer.}
\label{fig:6}
\end{figure}

\begin{figure}[htbp]
\epsfig{file=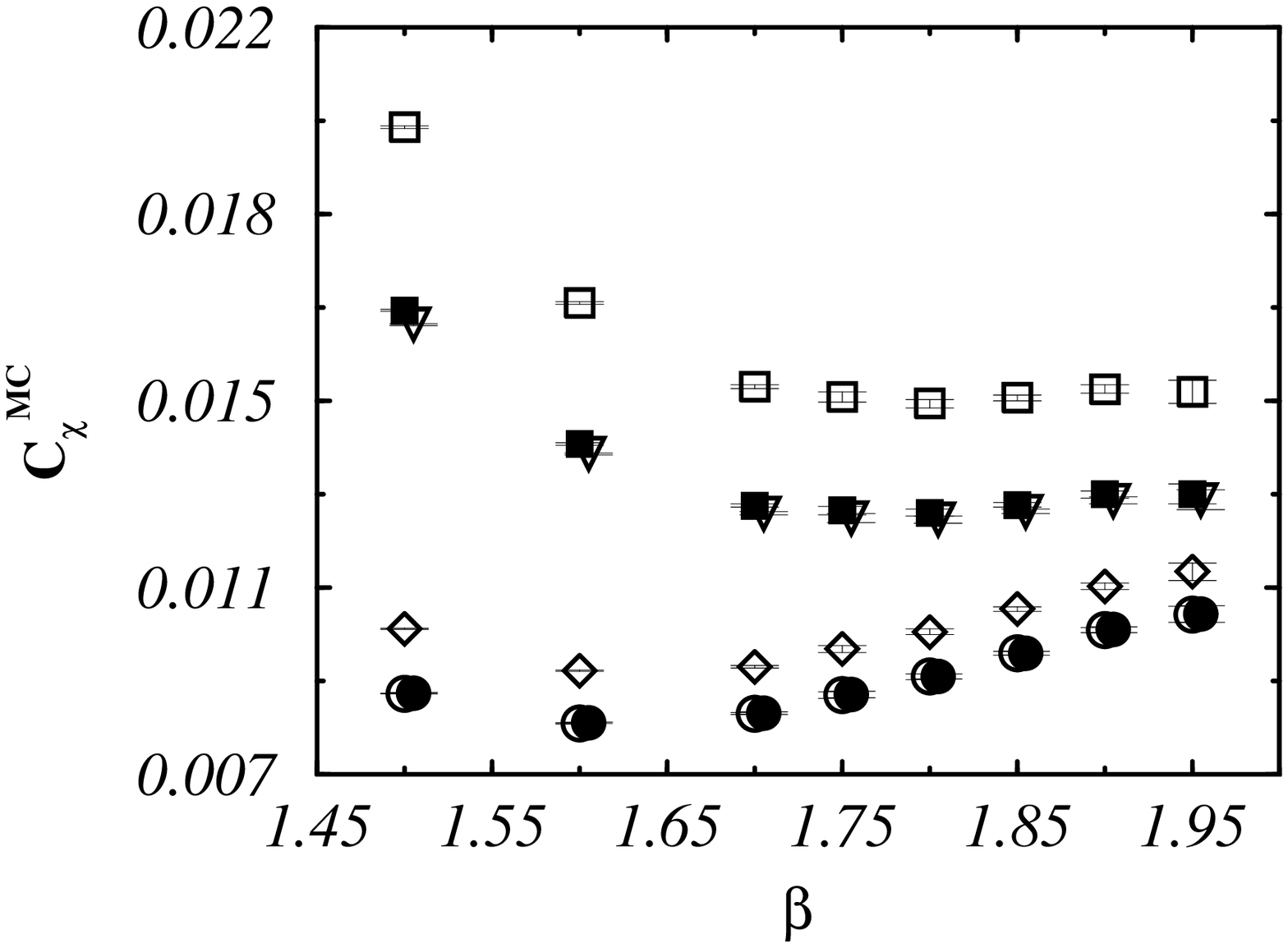,width=0.7\textwidth}
\caption{The non-perturbative constant
$C^{\rm MC}_{\chi}$ for the $O(3)$ model with standard action.
Monte Carlo data from \protect\cite{apos}. Open circles (full circles,
open diamonds) correspond to 2-loop (3-loop, 4-loop) in the
standard scheme; open squares (full squares, open triangles)
correspond to the 2-loop (3-loop, 4-loop) in the energy scheme.
Some data have been slightly shifted horizontally to render
the figure clearer.}
\label{fig:7}
\end{figure}

\begin{figure}[htbp]
\epsfig{file=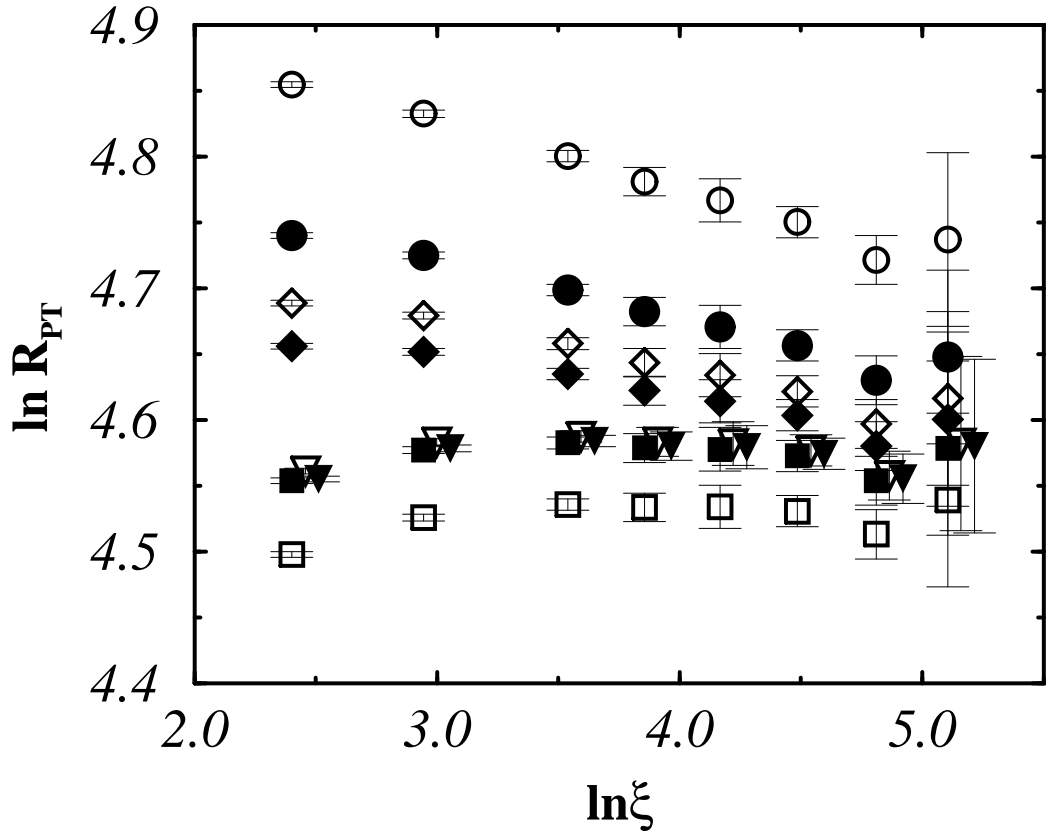,width=0.9\textwidth}
\caption{The $PT$ ratio 
for the $O(3)$ model with standard action.
Monte Carlo data from \protect\cite{apos}. Open circles (full circles,
open diamonds, full diamonds) correspond to 
1-loop (further corrections) in the
standard scheme; open squares 
(full squares, open triangles, full triangles)
correspond to the 1-loop (further corrections) in the energy scheme.
Some data have been slightly shifted horizontally to render 
the figure clearer.}
\label{fig:8}
\end{figure}

Finally we show the $PT$ ratio up to 4 loops for the standard
and energy schemes in Figure 8. The data for the standard scheme
is clearly not constant as already seen in \cite{patr2}. However
again the data in the energy scheme is particularly good and stable 
and allows the determination $\ln (C_\chi/C_\xi^2)= 4.57(2)$
in excellent agreement with the previous determination by using our
data for the Symanzik action (as it should this ratio is independent
of the regularization used). 

Our results for the $PT$ ratio and the magnetic susceptibility 
are 4.57(2) and 0.0130(5) respectively. These results, obtained by
using the standard action, agree with the previous ones extracted 
with the Symanzik action. Besides, the ${\cal O}(1/n^2)$ estimate of 
$C_\chi$~\cite{flyv} is in good accordance with our data.
The deviation from $C_\xi$ and the exact
result \cite{hase2} is still of the order $10\%$ even after the 
inclusion of the 4-loop correction in the energy scheme.

\subsection{The $O(8)$ model with standard action}

Our Monte Carlo data for the $O(8)$ model is shown in 
Table II. 
Our data agree with ref. \cite{wolf3} for the two values of
$\beta$ that we have in common. 
In Table VI we give the energy data taken from \cite{mendes}.
The perturbative expansion for the energy is reported in 
eq. (\ref{eq:n8}) of the appendix.

\begin{figure}[htbp]
\epsfig{file=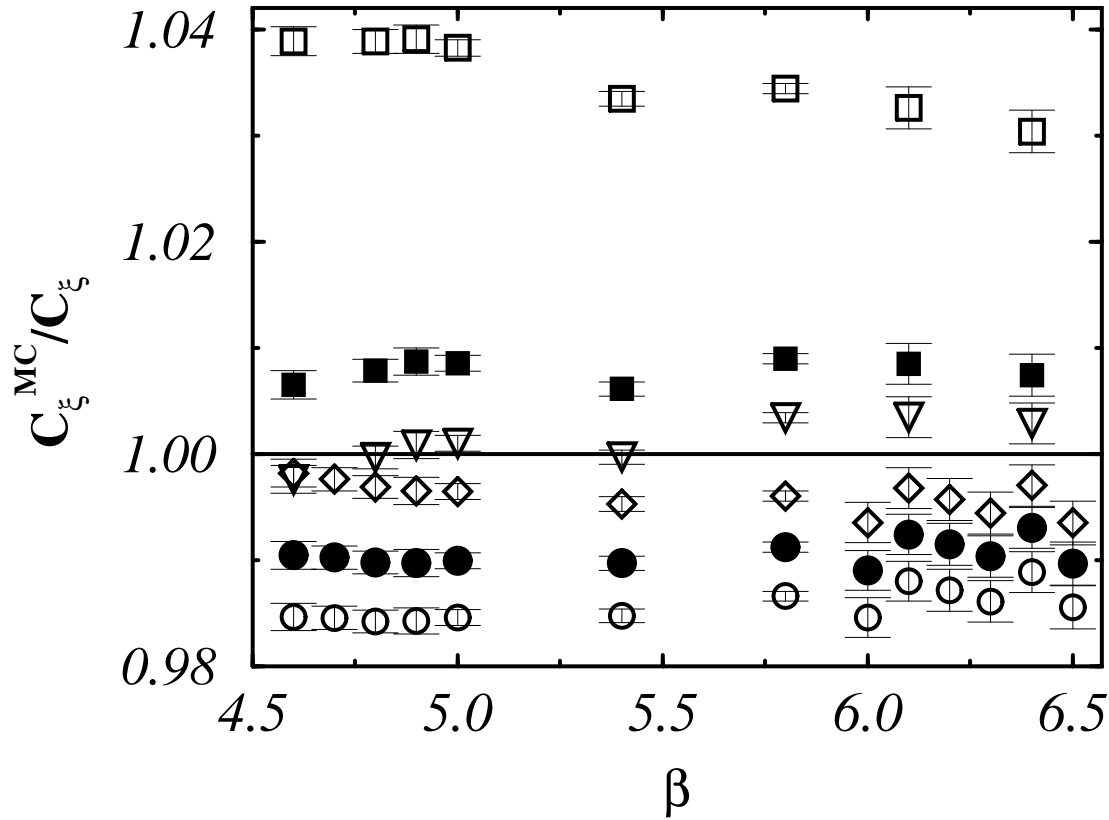,width=0.9\textwidth}
\caption{
The ratio 
$C^{\rm MC}_{\xi}/C_{\xi}$ for the $O(8)$ model with standard action.
Open circles (full circles, open diamonds) stand for the
2-loop (3-loop, 4-loop) approximation in the standard scheme.
Open squares (full squares, open triangles) stand for the
2-loop (3-loop, 4-loop) approximation in the energy scheme.}
\label{fig:9}
\end{figure}

In Figure 9 we show for the $O(8)$ model the equivalent of Figure 2.
The data converge towards 1 in both the standard and energy schemes.
The 4-loop energy scheme for the ratio $C^{\rm MC}_{\xi}/C_{\xi}$ yields
1 up to $\sim 0.5\%$.

\vskip 1cm

{\centerline {\bf Table VI}}
{\centerline {\it Energy data for the $O(8)$ model 
(from ref. \cite{mendes}).}}

\vskip 5mm

\moveright 2.3 in
\vbox{\offinterlineskip
\halign{\strut
\vrule \hfil\quad $#$ \hfil \quad & 
\vrule \hfil\quad $#$ \hfil \quad & 
\vrule \hfil\quad $#$ \hfil \quad \vrule \cr
\noalign{\hrule}
\beta & L & E  \cr
\noalign{\hrule}
4.6 & 128 & 0.603836(9)  \cr
\noalign{\hrule}
4.8 & 64 & 0.620987(10)  \cr
\noalign{\hrule}
4.9 & 128 & 0.629018(9)  \cr
\noalign{\hrule}
5.0 & 64 & 0.636812(10)  \cr
\noalign{\hrule}
5.4 & 64 & 0.664983(9)  \cr
\noalign{\hrule}
5.8 & 256 & 0.688885(3)  \cr
\noalign{\hrule}
6.1 & 256 & 0.704805(3)  \cr
\noalign{\hrule}
6.4 & 256 & 0.719168(2)  \cr
\noalign{\hrule}
}}

\vskip 5mm

 The figure clearly displays that the data approach 1 monotonically
as the number of loops increases. An important issue then
is to understand how big the successive corrections are. At leading
order in $1/n$ the coefficients $a_k$ in eq. (\ref{eq:scaling}) have
been computed up to $k$=8 \cite{cara2}. Comparing with the exactly
known coefficients $a_1$ and $a_2$ we see that the large-$n$
approximations $a_{1,2}(n=8)$ are correct up to 90\% and 60\%
respectively \cite{cara2,cara4}. Assuming a corrective factor
$f_k=1-2$ such that $a_k=f_k \cdot a_k(n=8)$ for all $k$, 
then one can see that the
next corrections are small and that the convergence towards 1 in
Figure 9 is meaningful.

 Figure 10 displays the magnetic susceptibility constant as extracted
from the Monte Carlo data, $C_{\chi}^{\rm MC}$. 
We do not show the data in the energy scheme at 2 loops as they are
very big $(\sim 0.108)$ and 
would expand too much the vertical scale of the figure.
Data tend to converge around the value $C_{\chi} \approx 0.102$.
Taking the results at 4-loop in the energy schemes we obtain
$C_\chi = 0.1028(2)$. The large-$n$ prediction is 
0.0915 (up to ${\cal O}(1/n)$, \cite{bisc}) and 0.103 (up to 
${\cal O}(1/n^2)$, \cite{flyv}). 
This ${\cal O}(1/n^2)$ estimate agrees with our result within
$\lesssim 0.5\%$ which is the same amount of 
deviation from unity 
seen in Figure 9 for the correlation length.
Therefore the $1/n$ expansion
agrees fairly well with our data.
Notice that data in the standard scheme do not converge monotonically;
indeed we have the sequence ``2-loop'' $>$ ``4-loop'' $>$ ``3-loop''.
This is due to the fact that the coefficients $b_1$ and $b_2$ in
eq. (\ref{eq:scaling}) have opposite signs.

\begin{figure}[htbp]
\epsfig{file=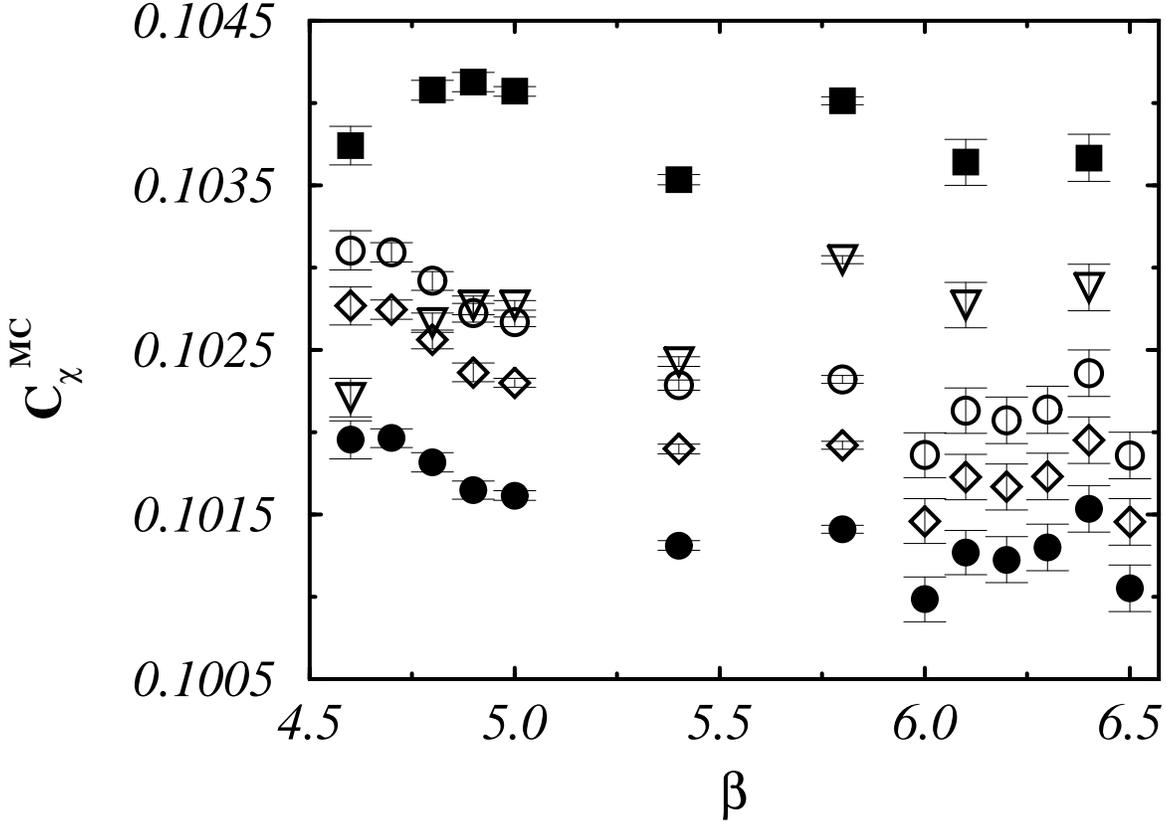,width=0.9\textwidth}
\caption{
The non-perturbative constant 
$C_{\chi}$ as extracted from the Monte Carlo data for the 
$O(8)$ model with standard action.
Open circles (full circles, open diamonds) stand for the
2-loop (3-loop, 4-loop) approximation in the standard scheme.
 Full squares (open triangles) stand for the
3-loop (4-loop) approximation in the energy scheme.}
\label{fig:10}
\end{figure}

\begin{figure}[htbp]
\epsfig{file=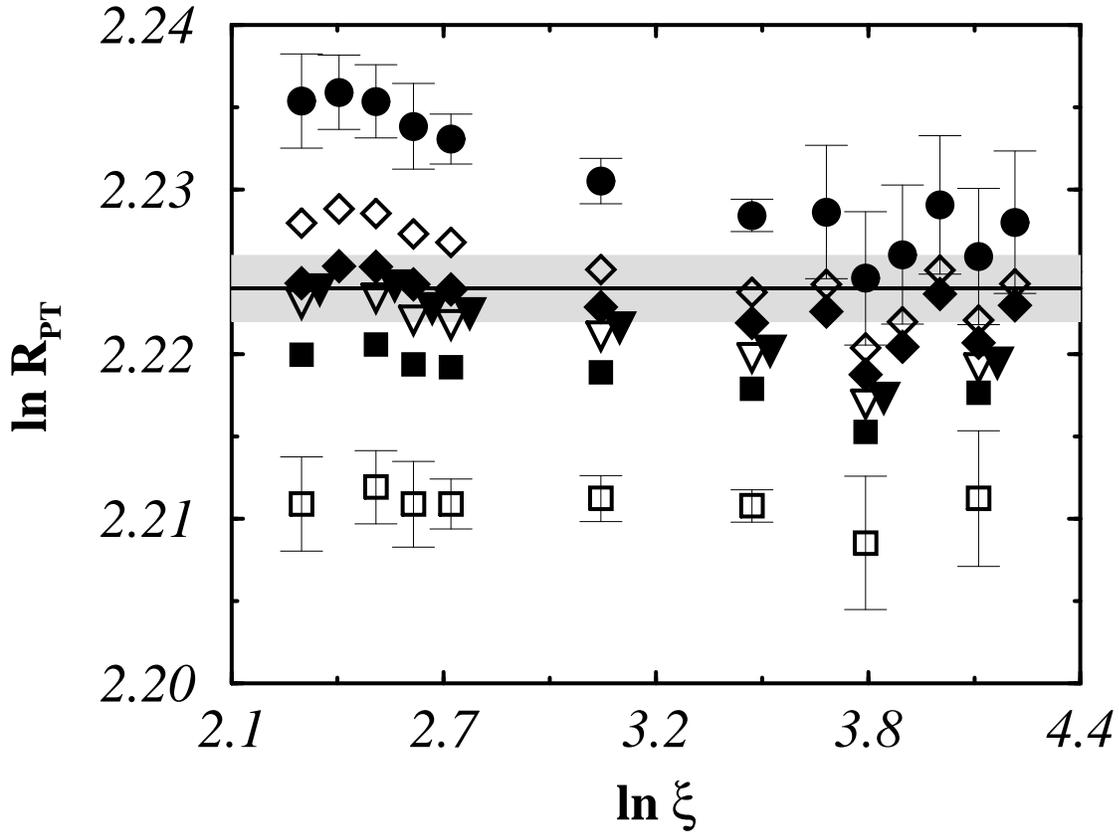,width=0.9\textwidth}
\caption{
The $PT$ ratio for the $O(8)$ model with 
standard action.
 Full circles (open diamonds, full diamonds) stand for the
2-loop (further loops) approximation in the standard scheme.
Open squares (full squares, open triangles, full triangles) stand for the
1-loop (further loops) approximation in the energy scheme.
The highest order corrections in the energy scheme have been slightly
shifted horizontally to render the figure clearer.}
\label{fig:11}
\end{figure}

 In Figure 11 we show the $PT$ ratio for the $O(8)$ model. We
show this ratio up to 4 loops. 
We do not show the data at 1 loop in the standard scheme because
again they lie far from the window shown in the vertical axis.
We have also omitted the error bars in the further corrections
to render the figure clearer.
The data stabilize for large
enough $\ln \xi$ after having included the non-universal corrections.
The convergence is extremely good. The straight horizontal line is
the prediction (and error) eq. (\ref{eq:ptratio}) taking the value of
eq. (\ref{eq:hasenfratz}) for $C_{\xi}$ and the result 0.1028(2)
for $C_{\chi}$ from the previous figure. Our data gives 
$R_{PT}=2.220(5)$.

\begin{figure}[htbp]
\epsfig{file=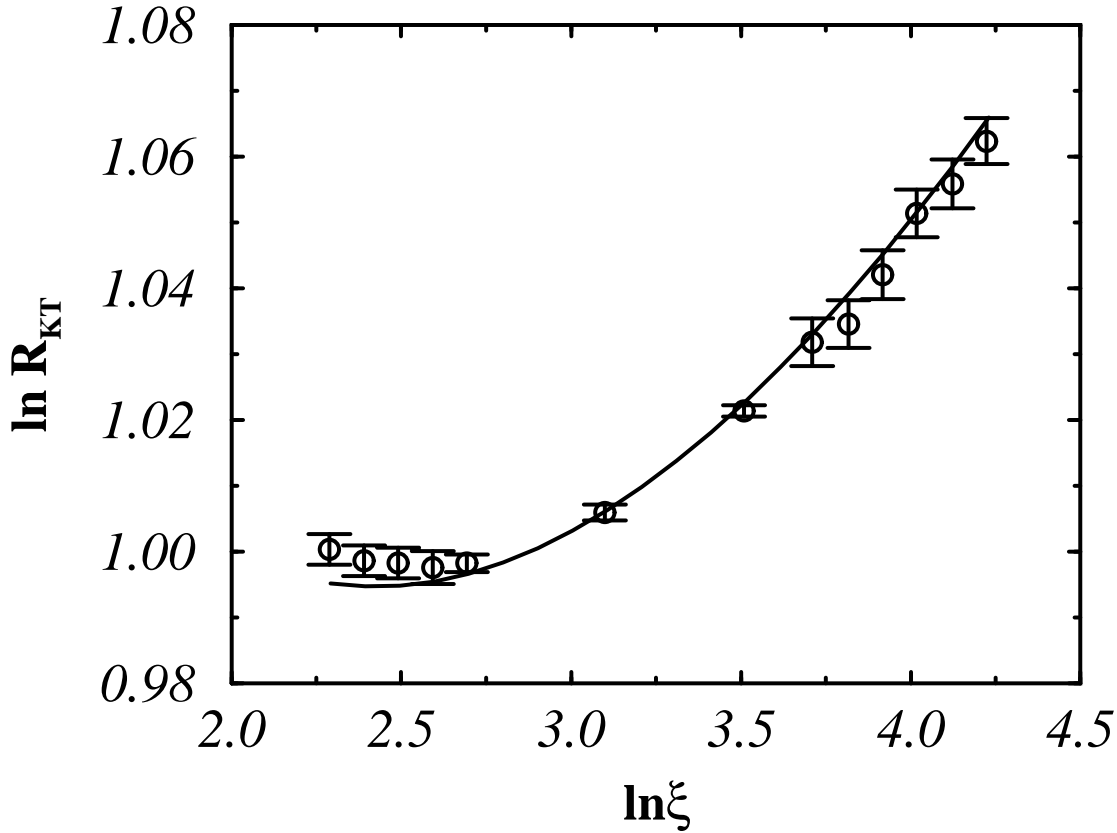,width=0.9\textwidth}
\caption{
The $KT$ ratio for the $O(8)$ model with 
standard action. The open circles are the data from our 
Monte Carlo simulation. The solid line is the prediction of
perturbation theory at 4 loops.}
\label{fig:12}
\end{figure}

 Finally Figure 12 shows the $KT$ ratio for the $O(8)$ model.
As in the $O(3)$ case, we have neglected the variation of $\tau^r$
inside the interval $4.6 < \beta < 6.5$. The solid line is the
$PT$ prediction for this ratio using eq. (\ref{eq:scaling}) up to
4 loops, eq. (\ref{eq:hasenfratz}) for $C_{\xi}$ and the
result from Figure 10 for 
$C_{\chi}$. We observe that the $KT$ ratio is not constant and that its
non-constancy is well explained by $PT$. 
Notice that the same set of values for $C_{\chi}$ and $C_\xi$ explain
well both the $PT$ and $KT$ ratios.

In conclusion, $PT$ works fairly well for the $O(8)$ model. The data
agree with the exact mass-gap \cite{hase2bis} with a precision about
0.5\%. Analogously the $1/n^2$ prediction for the magnetic susceptibility
is in fair accordance with our data within the same error. Moreover
the $PT$ and $KT$ ratios are well described by the $PT$ formulae
(\ref{eq:scaling}). 

\section{Conclusions}

We have done a Monte Carlo simulation for the $O(3)$ and $O(8)$
non-linear $\sigma$-models in 2 dimensions. 
The simulation was performed with the tree-level Symanzik action
for the $O(3)$ model 
and the standard action for the $O(8)$ model. We have improved the
statistics with respect to previous works and have taken advantage of
the recently calculated 4-loop corrections to scaling \cite{cara2}.
We have taken into account the systematic errors coming from the 
finite lattice size \cite{cara1} and from the different non-perturbative
constants for the correlation length \cite{cara2}. In order to
reduce them we made use of the correlation length data for 
the $\xi^{(2)}$ definition, eq. (\ref{eq:defxi}), and calculated
numerically the corrective factor to pass from $C_{\xi^{\rm exp}}$,
eq.~(\ref{eq:hasenfratz}), to $C_{\xi^{(2)}}$. The result for this
numerically calculated factor was in good agreement with the 
$1/n$ estimate (\ref{eq:constantscorr}). 

The ensemble of independent configurations 
was created with the fast Wolff algorithm,
\cite{wolf}. The independence of the measurements done on these
configurations was explicitly verified. However, in the small physical
size regime, $\rho=L/\xi \ll 1$ we discovered a worsening in the 
performance of
this algorithm. We argued that this fact can be explained by the presence
of large Fortuin-Kasteleyn clusters \cite{fk,sw} when $\rho \ll 1$, 
see Table III.

We have also made use of the data of ref. \cite{apos} for the $O(3)$
model with standard action.

In all cases we tested the perturbation theory predictions in both
the standard scheme (expansions in the bare coupling $1/\beta$) and
the energy scheme (energy modified coupling $1/\beta_E$). For this
purpose in the appendix we have computed the weak coupling expansion 
of the energy up to 4 loops for the standard action and 3 loops
for the Symanzik action. 
The fourth loop term in the standard action and the whole expansion
up to 3 loops in the Symanzik action are new results of the present
paper. Moreover, for the Symanzik action, we computed
two different operators, $E^S_1$ and $E^S_2$, in order to check the
validity of the energy scheme: they should give almost identical
results. This check was successful (see figures 2-4).

We saw that the results for the $O(3)$ model agree fairly well with
$PT$ in the energy scheme. The $PT$ ratio leads to an almost constant
already at 2 loops for both standard and Symanzik actions. 
This constant was
$\ln (C_\chi/C_\xi^2) =4.54(2)$ and 4.57(2) for the standard
and Symanzik actions respectively. 
The value observed for the non-perturbative constant $C_\xi$
differs from the prediction (\ref{eq:hasenfratz}) by almost 
$2-3\%$ in the Symanzik action and $\gtrsim 10\%$ for the standard one.
In both cases we refer to the results in the energy scheme. 
Even though these differences are still too large, they are much
smaller than when obtained from the expansion in the
standard $1/\beta$ ($\sim 20\%$).
The numbers for the constant $C_\chi$ are 0.0138(2) and 0.0130(5)
for the Symanzik and standard actions respectively (both in 
units of $\Lambda_{\rm standard}$). The $1/n^2$ prediction \cite{flyv}
is 0.0127. 
Besides, our determinations are in acceptable
accordance with the prediction \cite{balo,prepa}
$C_\chi \approx 0.0145$. This number has been extracted from the
non-perturbative constant $\lambda_1$ obtained in \cite{balo} and
the ratio between the on-shell and zero-momentum 
field-renormalization constants $q\equiv Z^{\rm zero-mom}/Z^{\rm on-shell}$
at large $\beta$.
This ratio is known in the $1/n$ expansion \cite{prepa} to be
$q=1+0.0132/n+\cdots$. We have computed this ratio from our Monte Carlo data,
$Z^{\rm zero-mom}$ being $\chi/\xi^{(2)}$ and 
$Z^{\rm on-shell}$ being the constant in front of the wall-wall
correlation function for large separation $t$
\begin{equation}
{\bar G}(t) \approx
 Z^{\rm on-shell} {{\exp\left(L/(2\xi^{\rm exp})\right)
} \over {L/\xi^{\rm exp}}} \cosh\left((t-L/2)/\xi^{\rm exp}\right).
\end{equation}
The value for $Z^{\rm on-shell}$ presented a plateau as
a function of $t$ in the interval
$\xi^{\rm exp}/2 \lesssim t \lesssim 3\xi^{\rm exp}/2$
and we chose the value at $t=\xi^{\rm exp}$.
In Table VII we give our numerical result for the ratio $q$ from our
data for the O(3) model with Symanzik action as a function of $\beta$. 
The average is $q=1.0035(18)$ which is in excellent agreement
with the $1/n$ expansion. The fact that $q$ is close to 1 up to 
few per cent, implies that the estimate $C_\chi \approx 0.0145$
is valid within few per cent. 
We see again a good performance of the $1/n$ expansion
even at $n=3$. In particular there is a considerable improvement from the
${\cal O}(1/n)$ approximation \cite{bisc} in eq. (\ref{eq:constantschi})
to the ${\cal O}(1/n^2)$ order \cite{flyv} in eq. (\ref{eq:flyvresults}). 
This fact makes us to suspect 
that also in eq.~(\ref{eq:constantscorr}) 
the ${\cal O}(1/n^2)$ term would notably improve the agreement 
with our numerical result for that ratio.

\vskip 1cm

{\centerline {\bf Table VII}}
{\centerline {\it Results for the ratio $q$ as a function of $\beta$.}

\vskip 5mm

\moveright -0.1 in
\vbox{\offinterlineskip
\halign{\strut
\vrule \hfil\quad $#$ \hfil \quad & 
\vrule \hfil\quad $#$ \hfil \quad & 
\vrule \hfil\quad $#$ \hfil \quad & 
\vrule \hfil\quad $#$ \hfil \quad & 
\vrule \hfil\quad $#$ \hfil \quad & 
\vrule \hfil\quad $#$ \hfil \quad & 
\vrule \hfil\quad $#$ \hfil \quad & 
\vrule \hfil\quad $#$ \hfil \quad \vrule \cr
\noalign{\hrule}
\beta & 1.40 & 1.45 & 1.50 & 1.55 & 1.60 & 1.65 & 1.70  \cr
\noalign{\hrule}
 q & 1.0052(40) & 1.0056(41) & 1.0028(46) & 1.0031(34) &
1.0007(89) & 1.0043(79) & 0.9981(60) \cr
\noalign{\hrule}
}}

\vskip 5mm

Recall that the Symanzik action has been designed to reduce 
lattice artifacts \cite{sym2}. However this
improvement can be overwhelmed by the large corrections to asymptotic
scaling. The effective schemes can cure this last problem. Hence the
combination of an improved action together with the 
use of an effective scheme should provide the best results. This may be
the reason for the good agreement between the $PT$ predictions and
our data from the $O(3)$ model with Symanzik action within the energy
schemes.

 Our analysis of the Monte Carlo results for the $O(8)$ model
reveals a satisfactory agreement between the $PT$ predictions and the
data. The value (\ref{eq:hasenfratz}) for $C_\xi$ is recovered 
within 0.5\% and the ${\cal O}(1/n^2)$ prediction for $C_\chi$ agrees
within less than 0.5\% with our result $0.1028(2)$,
(again there is a remarkable improvement between
the ${\cal O}(1/n)$ and ${\cal O}(1/n^2)$ calculations).
Analogously the $PT$ ratio tends to stabilize at
$\ln (C_\chi/C_\xi^2)=2.220(5)$; the same prediction
calculated from the previous value for 
$C_\chi$ and the exact $C_\xi$ (\ref{eq:hasenfratz})
is shown in Figure 11 as an horizontal line at 
$\ln R_{PT}=2.224(2)$.

 We have also checked the set of predictions of the $KT$ 
scenario for the $O(3)$ model with Symanzik action and the $O(8)$ 
model with standard action. Figures 5 and 12 show the results
for these two cases for the $KT$ ratio. 
None of them yield a constant as it happened
for the data of ref. \cite{patr2}. 
We stress the fact that our data have better resolution as the
error bars are almost one order of magnitude shorter than in \cite{patr2}.
As for the $O(3)$ model,
we showed that the probability of having a straight line after eliminating
the first two data points in Figure 5 is less than 10\%.
The situation for the $O(8)$ model is much clearer: the data
are definitively far from constant. In this case
perturbation theory predicts fairly well the trend of the data,
mainly for the largest correlations. It is worth noticing that 
the two ratios, $R_{KT}$ and $R_{PT}$, 
are well explained with the same set of
parameters $C_\xi$ and $C_\chi$ obtained from our analysis.
We could not draw a similar $PT$ prediction for the 
$R_{KT}$ ratio for the $O(3)$ model 
like the solid line in Figure 12 because the results for $C_\chi$ and
$C_\xi$ for the $O(3)$ model had less precision and the $KT$ ratio
is rather sensitive to the precision.

 We also tried a fit of the data for the correlation length 
to the $KT$ law (\ref{eq:ktxi}). The fit is unstable
because the actual value for $\beta_{KT}$ (if it is finite) 
is much larger than our working $\beta$'s

 In summary, $PT$ works well if one includes also the non-universal
corrections. Only the correlation length data for the $O(3)$ model
with standard action still stays far from the (\ref{eq:hasenfratz})
prediction, although these non-universal 
corrections improve the accordance by a factor
of 2. In this respect, we have seen that the energy scheme \cite{mart}
performs very well and it is a reliable scheme as explicitly 
proved by using two different operators with the
Symanzik data for $O(3)$. In ref. \cite{cara5} the authors calculate
the non-universal corrections to scaling for the spherical model,
discovering that they are absent for 
the energy scheme. We have seen that
this good behaviour is almost preserved at low values of $n$.

\section{Acknowledgements}

We thank Andrea Pelissetto for many useful and stimulating
conversations and for a critical reading of the manuscript 
and Paolo Rossi for a clarification about ref. \cite{balo}. 
B.A. also thanks Wolfhard Janke for a useful 
comment and Massimo Falcioni for correspondence about ref.
\cite{falc}. B.A. acknowledges financial support from an INFN contract.

\section{Electronic Memorandum} 

A progress report of the present 
work was sent to the Lattice-96 proceedings,
\cite{latt96}. While this progress report circulated as an hep-lat
preprint, a comment \cite{comment} appeared which motivated our reply
\cite{reply}.

\section{Appendix}

In this appendix we sketch the calculation of the energy 
up to 4 loops for the standard action and 3 loops for the
tree-level improved Symanzik action.

\subsection{Standard action}

We define the energy for the standard action as (not summed on $\mu$!)
\begin{equation}
E \equiv \langle {\vec \phi(0)} \cdot {\vec \phi(0+{\hat \mu})}  \rangle
\label{eq:energyst}
\end{equation}
which in the weak coupling expansion can be written as
\begin{equation}
E(\beta) = 1 - {w_1 \over \beta} -{w_2 \over \beta^2} -{w_3 \over \beta^3} -
{w_4 \over \beta^4} - \cdots
\label{eq:eexp}
\end{equation}
The first two coefficients $w_1$ and $w_2$ can be straightforwardly
computed giving
\begin{equation} 
w_1 = {{\left(n - 1\right)} \over 4}, \qquad \qquad \qquad
w_2 = {{\left(n - 1\right)} \over 32} .
\label{eq:w1w2}
\end{equation}

\leavevmode
\begin{center}
\begin{figure}[htbp]
\epsfig{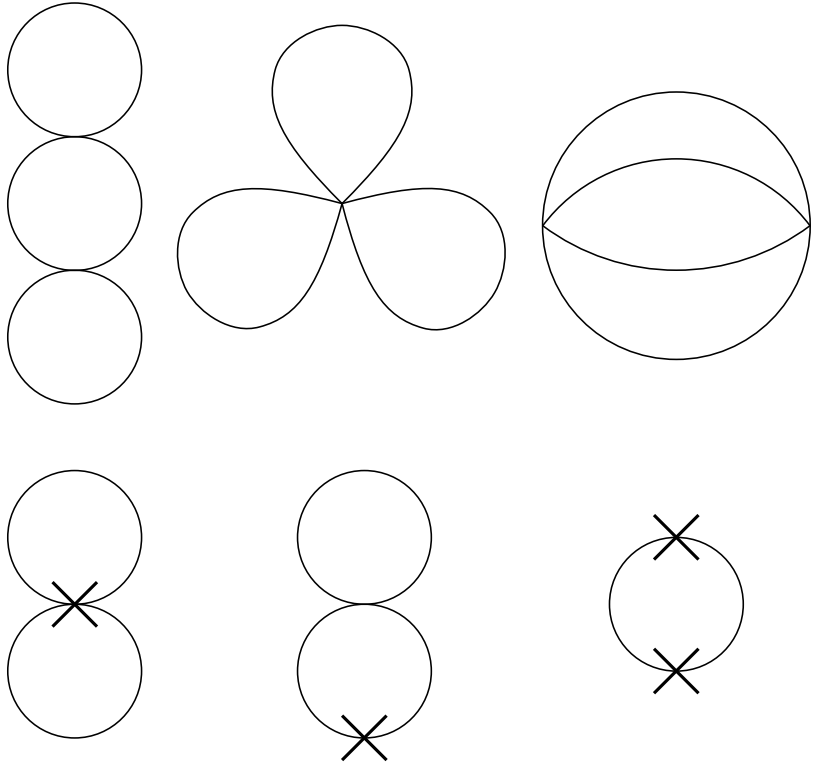}
\caption{
Feynman diagrams contributing to the 3-loop 
coefficient of the free energy. Crosses stand for insertions of
the measure lagrangian that comes from the Dirac delta in eq.
(\ref{eq:zeta}).}
\label{fig:13}
\end{figure}
\end{center}

\noindent The order ${\cal O}(1/\beta^3)$ coefficient has been computed in 
\cite{cara5} (for the $O(3)$ model it was also 
calculated in \cite{lusc} and for general $n$ in \cite{lulu}). 
We have checked their result
by computing the diagrams for the free energy in Figure 13 and by 
making use of the relationship
\begin{eqnarray}
E &=& {1 \over {2 \, V}} {{\partial} \over {\partial \beta}}
\ln Z, \nonumber \\
Z &\equiv& \int \, {\cal D}{\vec \phi}(x) \, \delta({\vec \phi}(x)^2 - 1)\,
\exp\left({-S^{\rm standard}}\right).
\label{eq:zeta}
\end{eqnarray}
$V$ is the space-time volume and ${\cal D}$ the standard functional
measure. 
In the evaluation of the Feynman diagrams the following identity is
useful
\begin{eqnarray}
{\widehat {\left( p_1 + p_2 \right)}}^2 &+&
{\widehat {\left( p_1 + p_3 \right)}}^2 +
{\widehat {\left( p_1 + p_4 \right)}}^2 
= {\hat p_1}^2 + {\hat p_2}^2 + {\hat p_3}^2 + {\hat p_4}^2 -
\Sigma_{1234}, \nonumber \\
&& \Sigma_{ijkl} \equiv
\sum_\mu {\hat p_{i\mu}} {\hat p_{j\mu}} {\hat p_{k\mu}} {\hat p_{l\mu}},
\label{eq:identity}
\end{eqnarray}
provided that $p_1+p_2+p_3+p_4=0 \;$ \cite{cara5}. We make use of the standard
notation, ${\hat p}_\mu \equiv 2 \sin(p_\mu/2)$ and ${\hat p}^2 \equiv
\sum_\mu {\hat p}_\mu^2$. Another relation useful during the evaluation of
tadpole diagrams is
\begin{equation}
{\widehat {\left( p_1 + p_2 \right)}}^2 = 
{\hat p_1}^2 + {\hat p_2}^2 - {1 \over 4} \, {\hat p_1}^2
\, {\hat p_2}^2 + \hbox{odd terms},
\label{eq:tadpole}
\end{equation}
valid for any pair of momenta $p_1$ and $p_2$.

The result for $w_3$ is
\begin{equation}
w_3 = {{\left(n - 1\right)^2} \over 16} \, K +
{{\left(n - 1\right)} \over 16} \, \left({1 \over 6} - K +
{1 \over 3} J\right).
\label{eq:w3}
\end{equation}
$K$ and $J$ are finite integrals 
\begin{eqnarray}
K &\equiv& \int^{+\pi}_{-\pi} 
{\rm D}_3 \;
{{ \Delta_{12} \Delta_{34} } \over
{ {\hat p_1}^2 \;{\hat p_2}^2 \;{\hat p_3}^2 \;{\hat p_4}^2}} =
0.0958876 \nonumber \\
J &\equiv& \int^{+\pi}_{-\pi} 
{\rm D}_3 \;
{{ \left(\Sigma_{1234}\right)^2} \over
{ {\hat p_1}^2 \;{\hat p_2}^2 \;{\hat p_3}^2\;{\hat p_4}^2}} =
0.136620
\label{eq:kj}
\end{eqnarray}
where the measure ${\rm D}_3$ is
\begin{equation}
{\rm D}_3 \equiv
{{{\rm d}^2p_1} \over {\left( 2 \pi \right)^2}}
{{{\rm d}^2p_2} \over {\left( 2 \pi \right)^2}}
{{{\rm d}^2p_3} \over {\left( 2 \pi \right)^2}}
{{{\rm d}^2p_4} \over {\left( 2 \pi \right)^2}}
\left( 2 \pi \right)^2 \delta(p_1+p_2+p_3+p_4) 
\label{eq:measure4}
\end{equation}
and 
\begin{eqnarray}
\Delta_{ij} &\equiv& 
{\widehat {\left( p_i + p_j \right)}}^2 -
{\hat p_i}^2 -{\hat p_j}^2 , \nonumber \\
\Delta_{i-j} &\equiv& 
{\widehat {\left( p_i - p_j \right)}}^2 -
{\hat p_i}^2 -{\hat p_j}^2 ,
\end{eqnarray}
$\Delta_{i-j}$ will be used later.

\leavevmode
\begin{center}
\begin{figure}[htbp]
\epsfig{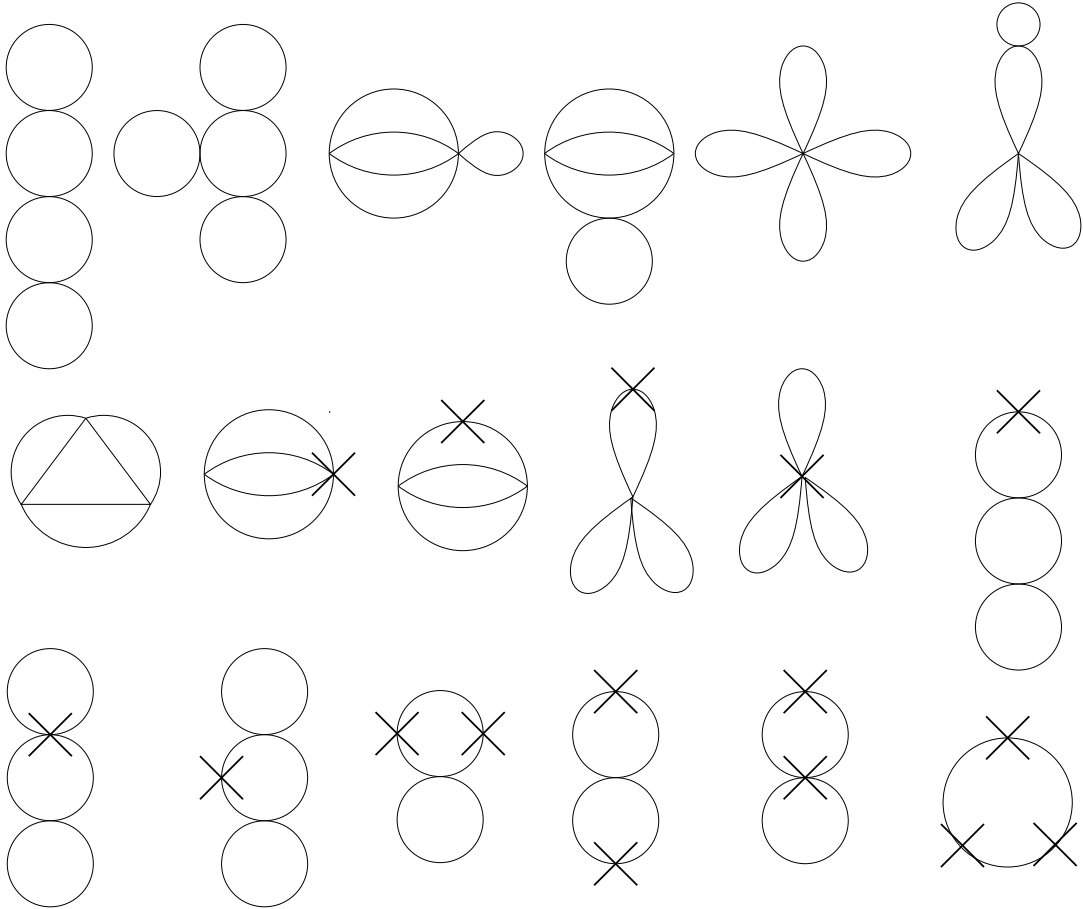}
\caption{
Feynman diagrams contributing to the 4-loop 
coefficient of the free energy. Same meaning as in Figure 9 
for the crosses.}
\label{fig:14}
\end{figure}
\end{center}

In Figure 14 we show the diagrams needed for the evaluation of $w_4$.
Again eqs. (\ref{eq:identity}) and (\ref{eq:tadpole}) are useful.
No new identities among momenta are needed.
The result is
\begin{eqnarray}
w_4 &=& {{3 \left(n - 1\right)} \over 8} \left(
{1 \over 128} - {1 \over 2} H_1 - {1 \over 4} H_2 - 
{1 \over 3} H_3 + {1 \over 24} J - 
{1 \over 8} K - {1 \over 4} H_5  \right) + \nonumber \\
&& 
{{3 \left(n - 1\right)^2} \over 8} \left(
{1 \over 256} + {1 \over 2} H_1 + {1 \over 4} H_2 +
{1 \over 3} H_3 + {1 \over 12} H_4 + 
{1 \over 8} K + {1 \over 3} H_5  \right) - \nonumber \\
&&
{{ \left(n - 1\right)^3} \over 32} H_5 .
\label{eq:w4}
\end{eqnarray}
$K$ and $J$ are given in eq. (\ref{eq:kj}) while $H_1$, ..., $H_5$ 
are genuine 4-loop integrals
\begin{eqnarray}
H_1 &\equiv& \int^{+\pi}_{-\pi} 
{\rm D}_4\;
{{ \Delta_{12} \;\Delta_{34} \; \Sigma_{1256}}  \over
{ {\hat p_1}^2 \;{\hat p_2}^2 \;{\hat p_3}^2 \;{\hat p_4}^2
\;{\hat p_5}^2\;{\hat p_6}^2}} =
0.0378134 \nonumber \\
H_2 &\equiv& \int^{+\pi}_{-\pi} 
{\rm D}_4\;
{{ \Delta_{34} \; \Sigma_{1234} \; \Sigma_{1256} }
 \over
{ {\hat p_1}^2 \;{\hat p_2}^2 \;{\hat p_3}^2 \;{\hat p_4}^2
\;{\hat p_5}^2\;{\hat p_6}^2}} =
-0.0322778 \nonumber \\
H_3 &\equiv& \int^{+\pi}_{-\pi} 
{\rm D}_4\;
{{ \Delta_{13} \; \Delta_{45} \; \Delta_{2-6}} \over
{ {\hat p_1}^2 \;{\hat p_2}^2 \;{\hat p_3}^2 \;{\hat p_4}^2
\;{\hat p_5}^2\;{\hat p_6}^2}} = -0.0136824
 \nonumber \\
H_4 &\equiv& \int^{+\pi}_{-\pi} 
{\rm D}_4\;
{{  \Sigma_{1234} \; \Sigma_{3456} \; \Sigma_{1256} }
 \over
{ {\hat p_1}^2 \;{\hat p_2}^2 \;{\hat p_3}^2 \;{\hat p_4}^2
\;{\hat p_5}^2\;{\hat p_6}^2}} =
0.0411085 \nonumber \\
H_5 &\equiv& \int^{+\pi}_{-\pi} 
{\rm D}_4\;
{{ \Delta_{12} \; \Delta_{34} \; \Delta_{56}
} \over
{ {\hat p_1}^2 \;{\hat p_2}^2 \;{\hat p_3}^2 \;{\hat p_4}^2
\;{\hat p_5}^2\;{\hat p_6}^2}} =
-0.0501528
\label{eq:h1h2h3uv}
\end{eqnarray}
The measure for the 4-loop integrals is
\begin{eqnarray}
{\rm D}_4 &\equiv&
{{{\rm d}^2p_1} \over {\left( 2 \pi \right)^2}}
{{{\rm d}^2p_2} \over {\left( 2 \pi \right)^2}}
{{{\rm d}^2p_3} \over {\left( 2 \pi \right)^2}}
{{{\rm d}^2p_4} \over {\left( 2 \pi \right)^2}}
{{{\rm d}^2p_5} \over {\left( 2 \pi \right)^2}}
{{{\rm d}^2p_6} \over {\left( 2 \pi \right)^2}} \times \nonumber \\
&& \qquad
\left( 2 \pi \right)^2 \delta(p_1+p_2+p_3+p_4) 
\left( 2 \pi \right)^2 \delta(p_5+p_6+p_3+p_4) .
\label{eq:measure6}
\end{eqnarray}

Numerically at 4 loops the expansion (\ref{eq:eexp}) reads
\begin{eqnarray}
E(\beta) = &1& - {{n-1} \over {4 \beta}} -
                 {{n-1} \over {32 \beta^2}} -
                 {{0.00726994\; (n-1) + 0.00599298\; (n-1)^2} \over \beta^3} -
\nonumber \\
           & & {{0.00291780\; (n-1) + 0.00332878\; (n-1)^2 + 
                 0.00156728\; (n-1)^3} \over \beta^4}.
\label{eq:nn}
\end{eqnarray}
 For $n=3$ and $n=8$ the expansion (\ref{eq:nn}) becomes
\begin{eqnarray}
E(\beta,\, n=3) &=& 1 - {1 \over {2 \beta}} - {1 \over {16 \beta^2}} -
 {0.03851 \over \beta^3} - {0.03169 \over \beta^4},
\label{eq:n3} \\
E(\beta,\, n=8) &=& 1 - {7 \over {4 \beta}} - {7 \over {32 \beta^2}} -
 {0.3445 \over \beta^3} - {0.7211 \over \beta^4}.
\label{eq:n8}
\end{eqnarray}

\subsection{Symanzik action}

 As for the Symanzik action, we have used two different local operators
to define the so-called energy-scheme (not summed over $\mu$!)
\begin{eqnarray}
E^S_1 &\equiv& 
 \langle {4 \over 3} {\vec \phi(0)} \cdot {\vec \phi(0+{\hat \mu})} 
- {1 \over 12} {\vec \phi(0)} \cdot {\vec \phi(0+2{\hat \mu})}  \rangle,
\label{eq:energysy1}
\\
E^S_2 &\equiv&
 \langle {\vec \phi(0)} \cdot {\vec \phi(0+{\hat \mu})}  \rangle.
\label{eq:energysy2}
\end{eqnarray}
The first operator is the energy density for the Symanzik-improved action,
hence its weak coupling expansion can be computed by evaluating the 
free energy and making use of eq. (\ref{eq:zeta}). In ref. \cite{digi}
it was computed up to 2 loops for the $n=3$ case. 
We have checked their result which for any $n$ can be written as
\begin{eqnarray}
E^S_1(\beta) &=& {15 \over 12} - 
{w^{S1}_1 \over \beta} -{w^{S1}_2 \over \beta^2} -
{w^{S1}_3 \over \beta^3} - \cdots ,\nonumber \\
&& w^{S1}_1 = { {\left( n - 1\right)} \over 4}, \nonumber \\
&& w^{S1}_2 = { {\left( n - 1\right)} \over 48}\; Y_1 \,
 \left( 1 - {5 \over 24} Y_1 \right) .
\label{eq:ws11ws12}
\end{eqnarray}
$Y_1$ is a 1-loop integral. The notation $\Pi_p$ will mean the
inverse propagator for the Symanzik action
\begin{equation}
\Pi_p \equiv {\hat p}^2 + {1 \over 12} \Box_p,
\qquad \qquad \qquad \Box_p \equiv \sum_\mu {\hat p}^4_\mu .
\end{equation}
The 1-loop integral is
\begin{equation}
Y_1 \equiv \int^{+\pi}_{-\pi} 
{{{\rm d}^2p} \over {\left( 2 \pi \right)^2}} \;
{{\Box_p} \over {\Pi_p}} = 2.043576 
\label{eq:y1}
\end{equation}

The 3-loop coefficient can be obtained by evaluating a set
of diagrams analogous to the one in Figure 13. Useful identities
are
\begin{eqnarray}
 \Pi_{p+q} &+&\Pi_{p+k} +\Pi_{p+r} =
\Pi_{p} +\Pi_{q} +\Pi_{k} +\Pi_{r} - \Sigma^S , \nonumber \\
&& \Sigma^S \equiv {4 \over 3}
\sum_\mu {\hat p_{\mu}} {\hat q_{\mu}} {\hat k_{\mu}} {\hat r_{\mu}}
- {1 \over 12}
\sum_\mu {\widehat {2p_{\mu}}} {\widehat {2q_{\mu}}} 
{\widehat {2k_{\mu}}} {\widehat {2r_{\mu}}} ,
\end{eqnarray}
valid whenever $p+q+k+r=0$ and
\begin{equation}
\Pi_{p+q} = \Pi_p + \Pi_q - {1 \over 12} \Pi_p \Box_q 
- {1 \over 12} \Pi_q \Box_p + {5 \over 144} \Box_p \Box_q 
+ \hbox{odd terms}
\end{equation}
for any pair of momenta $p$ and $q$.
The result for $w^{S1}_3$ is
\begin{eqnarray}
w^{S1}_3 &=& {{\left( n - 1 \right)^2} \over 16} K^S +
{{\left( n - 1 \right)} \over 2} \Bigg( {1 \over 24} + {1 \over 24} J^S
- {1 \over 8} K^S + {1 \over 288} Y_2 - \nonumber \\
&& Y_1 \left( {5 \over 96} + {5 \over 1728} Y_2 \right) +
Y_1^2 \left( {11 \over 384} + {25 \over 41472} Y_2 \right) -
{205 \over 41472} Y_1^3 \Bigg).
\label{eq:ws13}
\end{eqnarray}
$Y_2$ is a 1-loop integral
\begin{equation}
Y_2 \equiv \int^{+\pi}_{-\pi} 
{{{\rm d}^2p} \over {\left( 2 \pi \right)^2}} \;
{{\Box_p^2} \over {\left(\Pi_p\right)^2}} = 4.783071
\label{eq:y2}
\end{equation}
The 3-loop integrals are
\begin{eqnarray}
K^S &\equiv& \int^{+\pi}_{-\pi} 
{\rm D}_3 \;
{{ \Delta^S_{12} \Delta^S_{34} } \over
{ {\Pi_{p_1}} \;{\Pi_{p_2}} \;{\Pi_{p_3}} \;{\Pi_{p_4}}}} =
0.0673316 \nonumber \\
J^S &\equiv& \int^{+\pi}_{-\pi} 
{\rm D}_3 \;
{{ \left(\Sigma^S\right)^2} \over
{ {\Pi_{p_1}} \;{\Pi_{p_2}} \;{\Pi_{p_3}} \;{\Pi_{p_4}}}} =
0.104551
\label{eq:ksjs}
\end{eqnarray}
where $\Delta^S_{12} \equiv \left( \Pi_{p1+p2} -
\Pi_{p1} - \Pi_{p2}\right)$.

 Numerically $E^S_1$ is
\begin{eqnarray}
E^S_1 (\beta) =&&{15 \over 12} - {{n-1} \over {4 \beta}} -
                    {{0.0244486\;(n-1)} \over { \beta^2}} - \nonumber \\
               &&   {{0.00449054\; (n-1) + 0.00420822\; (n-1)^2} \over
                     \beta^3}.
\label{eq:es1nn}
\end{eqnarray}
 For $n=3$ it is
\begin{equation}
E^S_1(\beta, n=3) = {15 \over 12}
- {1\over {2 \beta}} - {0.04890 \over \beta^2} -
{0.02581 \over \beta^3}.
\label{eq:es1num}
\end{equation}

\leavevmode
\begin{center}
\begin{figure}[htbp]
\epsfig{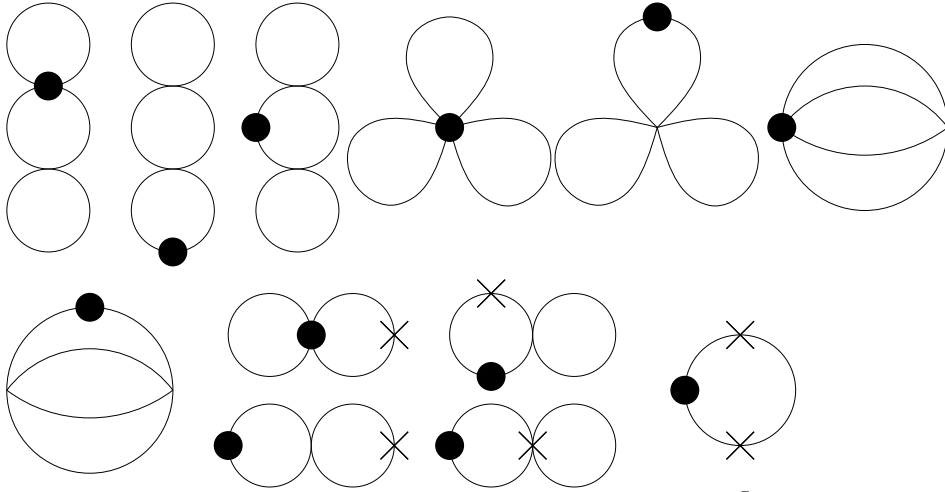}
\caption{Feynman diagrams contributing to the 3-loop 
coefficient of the operator $E^S_2$, eq. (\ref{eq:energysy2}). 
Crosses and black spots stand for insertions of measure and 
the operator $E^S_2$ respectively.}
\label{fig:15}
\end{figure}
\end{center}

 The second operator used is eq. (\ref{eq:energysy2}). 
The computation of the coefficients in the weak expansion 
\begin{equation}
E^S_2(\beta) = 1 - {w^{S2}_1 \over \beta} -{w^{S2}_2 \over \beta^2} -
{w^{S2}_3 \over \beta^3} - \cdots 
\end{equation}
requires the evaluation of diagrams with an
insertion of the operator in (\ref{eq:energysy2}). In Figure 15
we show the diagrams necessary for the 3-loop coefficient.
The results for all coefficients are
\begin{eqnarray}
w^{S2}_1 &=& {{\left(n - 1\right)} \over 4} \left( 1- {1\over 12}
Y_1 \right) ,\nonumber \\
w^{S2}_2 &=& {{\left(n - 1\right)} \over 32} \left( Y_1 
\left( {3 \over 2} + {5 \over 216} Y_2 \right) - {1 \over 18}
Y_2 - {49 \over 144} Y_1^2 -1 \right) \nonumber \\
w^{S2}_3 &=& {{\left(n - 1\right)^2} \over 16} \left( 2 
{\overline {K^S} } - {\widetilde K^S} \right) +
{{\left(n - 1\right)} \over 16} \Bigg(
{\widetilde K^S} - 2 {\overline {K^S} } + {2 \over 3} 
{\overline {J^S} } - {1 \over 3} {\widetilde {J^S} } + ,\nonumber \\
&& {13 \over 24} - {3529 \over 41472} Y_1^3 + {23 \over 288} Y_2 +
{5 \over 5184} Y_2^2 -  {1 \over 432} Y_3 + \nonumber \\
&& Y_1^2 \left( {61 \over 128} + {695 \over 41472} Y_2 - 
{25 \over 62208} Y_3 \right) + \nonumber \\
&& Y_1 \left( -{27 \over 32} - {127 \over 1728} Y_2 -
{25 \over 62208} Y_2^2 + {5 \over 2592} Y_3 \right) \Bigg).
\label{ws21ws22ws23}
\end{eqnarray}

The integrals are
\begin{eqnarray}
Y_3 &\equiv& \int^{+\pi}_{-\pi} 
{{{\rm d}^2p} \over {\left( 2 \pi \right)^2}} \;
{{\Box_p^3} \over {\left(\Pi_p\right)^3}} = 11.816615 \nonumber \\
{\widetilde {K^S}} &\equiv& \int^{+\pi}_{-\pi} {\rm D}_3 \;
{{ \Delta^S_{12} \; \Delta_{34} } \over {
{ {\Pi_{p_1}} \;{\Pi_{p_2}} \;{\Pi_{p_3}} \;{\Pi_{p_4}}}}} =
0.0578002 \nonumber \\
{\overline {K^S}} &\equiv& \int^{+\pi}_{-\pi} {\rm D}_3 \;
{{ \Delta^S_{12} \; \Delta^S_{34} \; {\hat p_1}^2} \over
 { \left({\Pi_{p_1}}\right)^2 \;{\Pi_{p_2}} \;{\Pi_{p_3}} \;{\Pi_{p_4}}}} =
0.0572726 \nonumber \\
{\widetilde {J^S}} &\equiv& \int^{+\pi}_{-\pi} {\rm D}_3 \;
{{ \Sigma_{1234} \; \Sigma^S } \over
{ {\Pi_{p_1}} \;{\Pi_{p_2}} \;{\Pi_{p_3}} \;{\Pi_{p_4}}}} =
0.0809553 \nonumber \\
{\overline {J^S}} &\equiv& \int^{+\pi}_{-\pi} {\rm D}_3 \;
{{ \Sigma^S \; \Sigma^S\; {\hat p_1}^2} \over
{ \left({\Pi_{p_1}}\right)^2 \;{\Pi_{p_2}} \;{\Pi_{p_3}} \;{\Pi_{p_4}}}} =
0.0867806
\end{eqnarray}

 Numerically $E^S_2$ is
\begin{eqnarray}
E^S_2 (\beta) = &1& - {{0.207425\; (n-1)} \over { \beta}} -
                    {{0.0189010\;(n-1)} \over { \beta^2}} - \nonumber \\
                &&  {{0.00353381\; (n-1) + 0.00354656\; (n-1)^2} \over
                     \beta^3}.
\label{eq:es2nn}
\end{eqnarray}
 For $n=3$ it is
\begin{equation}
E^S_2(\beta, n=3) = 1- {0.41485 \over \beta} - {0.03780 \over \beta^2} -
{0.02125 \over \beta^3} .
\label{eq:es2num}
\end{equation}

Another method to calculate the previous coefficients has been 
proposed in \cite{o3,su2,becc}.
The Monte Carlo determination of any
operator at large $\beta$ can be straightforwardly compared to
its perturbative expansion, allowing an estimate of the 
perturbative coefficients. In the last three rows of Table IV we
give the values of $E^S_1$ and $E^S_2$ for $\beta =5$, 10, 15.

The ${\cal O}(1/\beta)$ coefficient can be obtained comparing the
energy at $\beta=15$ with the expression $15/12-w^{S1}_1/\beta$
and $1-w^{S2}_1/\beta$.
We obtain $w^{S1}_1=0.502(1)$ and $w^{S2}_1=0.4167(3)$. 

Assuming that the exact first order coefficient is known,
one can use the value of the energy at
$\beta=10$ to determine the ${\cal O}(1/\beta^2)$ coefficient
obtaining $w^{S1}_2=0.051(2)$ and $w^{S2}_2=0.039(4)$.

Similarly, by using the exact two first coefficients and the value
at $\beta=5$ one obtains $w^{S1}_3=0.028(2)$ and $w^{S2}_3=0.022(5)$.

These results are clearly influenced by the next orders
and likely also by the small size ($L=100$) of the lattice used to 
calculate the energies for these large $\beta$'s.
A better analysis must use a global fit for all coefficients and
higher precision in the Monte Carlo determination of the operator.
Here we have used this technique just as an approximate check for our 
analytical computation.


\end{document}